# Control of Giant Topological Magnetic Moment and Valley Splitting in Trilayer Graphene

Zhehao Ge[1,*], Sergey Slizovskiy[2,3,*], Frédéric Joucken[1], Eberth A. Quezada[1], Takashi Taniguchi[4], Kenji Watanabe[5], Vladimir I. Fal'ko[2,3,6,†], Jairo Velasco Jr.[1, †]

[1]*Department of Physics, University of California, Santa Cruz, California, USA*

[2]*Department of Physics and Astronomy, University of Manchester, Oxford Road, Manchester, M13 9PL, UK*

[3]*National Graphene Institute, University of Manchester, Booth Street East, Manchester, M13 9PL UK*

[4]*International Center for Materials Nanoarchitectronics National Institute for Materials Science, 1-1 Namiki, Tsukuba, 305-0044, Japan*

[5] *Research Center for Functional Materials National Institute for Materials Science, 1-1 Namiki, Tsukuba, 305-0044, Japan*

[6] *Henry Royce Institute for Advanced Materials, Manchester, M13 9PL, UK*

[*]These authors contributed equally to this manuscript.
[†]Email: jvelasc5@ucsc.edu, Vladimir.Falko@manchester.ac.uk

# Abstract:

Bloch states of electrons in honeycomb two-dimensional crystals with multi-valley band structure and broken inversion symmetry have orbital magnetic moments of a topological nature. In crystals with two degenerate valleys, a perpendicular magnetic field lifts the valley degeneracy via a Zeeman effect due to these magnetic moments, leading to magnetoelectric effects which can be leveraged for creating valleytronic devices. In this work, we demonstrate that trilayer graphene with Bernal stacking (ABA TLG), hosts topological magnetic moments with a large and widely tunable valley g-factor ($g_v$), reaching a value $g_v \sim 1050$ at the extreme of the studied parametric range. The reported experiment consists in sublattice-resolved scanning tunneling spectroscopy under perpendicular electric and magnetic fields that control the TLG bands. The tunneling spectra agree very well with the results of theoretical modelling that includes the full details of the TLG tight-binding model and accounts for a quantum-dot-like potential profile formed electrostatically under the scanning tunneling microscope tip.

The orbital magnetic moment stemming from the rotational motion of electrons is ubiquitous in nature. It can be found in a variety of systems from single atoms to complex crystals, and can influence the magnetic properties of these systems. In recent years, topological magnetic moments emerging from self-rotating wave packets[1] have been discovered in 2D Van der Waals crystals with broken inversion symmetry.[2-6] Experimental manifestations of the topological magnetic moments have been observed lately, including the valley Zeeman effect,[2-17] spontaneous orbital ferromagnetism,[18, 19] and orbital magnetoelectric effects.[20-22] The former is important for valleytronics because it enables control of individual valley states, while the latter two could potentially facilitate new ultra-low power magnetic devices. To harness the valley Zeeman and orbital magnetoelectric effects in 2D crystals, systems with topological magnetic moments both large and tunable via gate modulation are desirable. The possibility to achieve these properties have been separately demonstrated with Bernal stacked bilayer graphene (BLG), offering[16] a tunable valley g-factor ($g_v$)$\sim 40 - 120$, and moiré superlattices in graphene, with[4] large $g_v \sim 2500$.

Here we realize a giant gate-tunable topological magnetic moment in naturally occurring Bernal stacked trilayer graphene (ABA TLG) by utilizing its peculiar band structure. Due to the mirror symmetry of ABA TLG (Fig. 1a), its electronic spectra can be viewed as overlapping bilayer graphene (BLG) and weakly gapped monolayer (MLG) bands.[23] A full tight-binding calculation of the ABA TLG band structure in the absence of a perpendicular electric field is plotted in Fig. 1b, where the effective MLG and BLG bands (both gapped) are indicated by the blue cones and semi-transparent red shells. The gaps and mutual alignment of the two bands are tunable by the encapsulation environment, gating, and doping. This feature offers an opportunity to engage states with a large topological magnetic moment and therefore giant $g_v$ specific to weakly gapped monolayers.[1,4,6] This is because with a similar gap size, gapped MLG has a much larger orbital magnetic moment compared to gapped BLG (see further discussion in supporting information (SI) S12).

In this work we use scanning tunneling microscopy/spectroscopy (STM/STS) to measure this giant $g_v$ and study the tunable topological magnetic moments of the effective MLG band in ABA TLG. The ABA TLG and hexagonal boron nitride (hBN) heterostructure for our STS study is fabricated with a conventional polymer-based transfer method[24] (see SI section S1 for sample fabrication details). ABA TLG and hBN are misaligned intentionally to avoid any spectral reconstruction near the charge neutrality point (CNP)[25], which is the energetic region of interest in our study. The measurement setup for our experiments is shown in the upper panel of Fig. 2a. The STM tip is grounded, and a bias voltage $V_S$ is applied between the STM tip and ABA TLG to induce a tunneling current. In addition, a backgate voltage $V_G$ is applied between the doped silicon and ABA TLG to institute an out-of-plane electric field that shifts the TLG Fermi energy and modifies the TLG band structure.[26] To avoid influence from adsorbates we performed all STS measurements at the centers of atomically pristine regions that were no smaller than $20 \times 20$ nm$^2$ (see typical topography of such region in SI section S1). The lower panel of Fig. 2a shows a typical topography at the center of such a region where the tunneling spectra were acquired. A clear triangular lattice is visible, which agrees with prior STM studies of ABA TLG supported on metals and SiC.[27,28] Furthermore, no moiré pattern is observed in our topography scans, thus indicating the ABA TLG and hBN are indeed misaligned.

A model atomic structure is overlaid on top of the measured topography in Fig. 2a that indicates the ABA TLG sublattices (for sublattice identification method see in SI section S3). The grey and bright spots correspond to sublattices $A_1$ and $B_1$, respectively. Both of these sublattices reside on the top layer, as shown in Fig. 1a. In contrast, the dark spot corresponds to sublattice $A_2$, which resides on the middle layer. Since STM is mostly sensitive to surface states, we expect the tunneling signal from our measurements to consist primarily of contributions from the top ABA TLG layer, hence sublattices $A_1$ and $B_1$ will dominate our STS measurements.

Typical gate resolved STS results for sublattices $A_1$ and $B_1$ are shown in Figs. 2b and 2c, respectively. To reduce the influence of slight deviations from the target sublattice for a single measurement, the tunneling spectra at each gate voltage shown in Figs. 2b and 2c correspond to an average of spectra at nine different targeted locations (see SI section S4 for the STS results before averaging). Interestingly, the spectra for sublattice $A_1$ exhibit a prominent $dI/dV_S$ peak (marked by a black dot) that diminishes in intensity and shifts toward the positive bias voltage with decreasing $V_G$. We find the strong $dI/dV_S$ peak is only present on sublattice $A_1$ (Figs. 2b,c). Notably, this feature was absent in previous gate resolved STS studies of ABA TLG.[29, 30]

Intrigued by this finding we next performed gate and sublattice resolved STS on the ABA TLG/hBN heterostructure in finite and out of plane magnetic field $B$. Our aim was to investigate the possibility of valley splitting in this system. Figure 3a shows the experimentally measured tunneling spectra on sublattice $A_1$ at $V_G = 30$ V with different $B$. The most prominent feature in these data is the strong $dI/dV_S$ peak that splits into two as $B$ is increased. This behavior was also observed at different $V_G$ on sublattice $A_1$ but not on sublattice $B_1$ (see SI section S5 for additional data). In addition, we found lower intensity satellite $dI/dV_S$ peaks emerge on the positive $V_s$ side as $B$ is increased. In contrast to the prominent sublattice dependent peaks, these satellite $dI/dV_S$ peaks were observed on both sublattices and at different $V_G$. Figure 3b shows the dependence of the peak splitting energy $\Delta E$ on $B$ at $V_G = 30$ V as red dots, we find the relationship between $\Delta E$ and $B$ is not linear. This nonlinear behavior is also observed at different $V_G$ (see SI section S11 for additional data).

The emergent $dI/dV_S$ peak observed on sublattice $A_1$ and its splitting in $B$ can be understood as resulting from a gapped MLG quantum dot (QD) with large topological magnetic moments. As shown in Fig. 1a, the antisymmetric wavefunction combination of sublattices $A_1$ and $A_3$ (blue shading) and $B_1$ and $B_3$ (orange shading) can be mapped onto a new sublattice B and A of an effective MLG lattice that gives rise to effective MLG bands.[26] Because of the $\gamma_2$ and $\gamma_5$ hopping energy difference and the onsite energy difference between the trimer and non-trimer sites ($\Delta_{AB}$), the effective MLG sublattices have different energies (broken inversion symmetry), leading to a light-mass Dirac spectrum with large topological magnetic moments.

Importantly, due to the capacitive coupling between the STM tip and ABA TLG, a shallow and smooth positive potential well is induced in ABA TLG, yielding an electrostatically defined QD.[31, 32] As depicted in the lower left panel of Fig. 3c, the positive potential well induced by the STM tip raises the energy of valence band MLG states into the bandgap, making these states localized and forming a valley degenerate QD state. This emerging QD state can explain the strong $dI/dV_S$ peak on sublattice $A_1$ (Fig. 2b) where the MLG states near the valence band edge reside. A comparison between the calculated LDOS for ABA TLG with and without a tip potential well can be found in SI section S8. Importantly, the localized state assists experimental detection of valley splitting in low $B$ (see discussion in SI section S9). Furthermore, with increasing $V_G$, the gap size ($\Delta$) of the effective MLG band increases (see SI section S6 for $\Delta$ determination details). This leads to an enhanced quantum confinement at higher $V_G$, which explains the increasing $dI/dV_S$ peak height at higher $V_G$ in Fig. 2b.

By applying an out of plane *B*, the valley degeneracy of the effective MLG QD state is lifted, thus explaining the splitting of the observed peak in *B*. As schematized in Fig. 1c, the topological magnetic moments $M_z(\vec{k}) = \tau \frac{e}{\hbar} \frac{\Delta}{[\Delta/(\hbar v_F)]^2 + 4|\vec{k}|^2}$ ($v_F$ is the Fermi velocity of the MLG bands, $\tau = +1$ and $-1$ for K' and K valley, respectively) of the effective MLG bands in K and K' valleys are both out of plane and with opposite orientations. Thus, an out of plane *B* will couple to the opposite $\vec{M} = \hat{z}\tau M_z$ of the electrons in the two valleys and generate valley splitting, as schematized in the lower right panel of Fig. 3c. Using this simple picture, $\Delta E$ can be approximated as $2|\vec{M} \cdot \vec{B}|$, which can also be expressed as $g_v \mu_B B$. Here $\mu_B$ is the Bohr magneton, and $g_v$ is defined as the valley g factor. With increasing *B*, the magnetic field confinement starts to dominate over the QD localization, and the valley splitting is expected to gradually start following the splitting between the Landau level (LL) 0- and LL1-, which is nonlinear as plotted by the green line in Fig. 3b. To fully account for the observed non-linearity, we further consider the influence of the tip potential on LL0- and LL1- (red line in Fig. 3b) as well as the effect from the MLG/BLG band mixing induced by a vertical electric field (orange line in Fig. 3b). After incorporating these additional effects in our theory, the predicted $\Delta E$ (blue line in Fig. 3b) shows good agreement with the experimentally extracted $\Delta E$ value in Fig. 3b (a more detailed discussion can be found in SI section S11).

Having understood the observed valley splitting in *B* for ABA TLG, we now discuss its gate tunability. The gap of the effective MLG band depends on the out-of-plane electric field ($E_z$), which can be expressed as $\Delta = \frac{1}{2}\sqrt{\gamma_2^2 + (U_1 - U_3)^2} + \frac{\gamma_5}{2} - \Delta_{AB}$. Here $U_1 - U_3 \propto E_z$ is the interlayer energy difference between the top and bottom layer of ABA TLG. Modulation of this quantity by $V_G$ controls the intensity of the inversion symmetry breaking in the top TLG layer, which leads to a gate tunable MLG gap ($\Delta$). Importantly, this tunable $\Delta$ will give rise to tunable topological magnetic moments in MLG bands. As shown in Fig. 1c, by increasing Δ from 14 meV to 26 meV, the maximum value of the topological magnetic moment changes from $808\mu_B$ to $442\mu_B$. Such gate tunable topological magnetic moments also yield gate tunable $g_v$.

To study the gate tunable $g_v$ in ABA TLG, we performed $dI/dV_S(V_S, B)$ measurements at different $V_G$. We first compare the experimental result with a simulation based on a full ABA TLG tight binding (TB) model with a potential well (details can be found in SI section S7). Figure 4a shows the second derivative of a measured $dI/dV_S(V_S, B)$ with high $B$ resolution, for which $V_G = 30$ V, the red features correspond to $dI/dV_S$ peaks. The STM tip used to acquire these data lack clear atomic resolution, as a result, we expect the tunneling spectra in this dataset is a mixture of the states from sublattices $A_1$ and $B_1$. Additional sublattice resolved $dI/dV_S(V_S, B)$ color plots measured at a single $V_G$ from a different STM tip can be found in SI section S14. Figure 4b is a simulated $\partial^2 LDOS/\partial E^2(E, B)$ at $V_G = 30$ V, the $LDOS$ from sublattice $A_1$ and $B_1$ are mixed together with a ratio of 10:1 to better reflect the nature of the STM tip used for the associated measurements. The experiment and simulation display good qualitative agreement, at high $B$ they both show a splitting peak and LLs below and above the Fermi level, respectively. Quantitative differences between the experiment and simulation can be attributed to parameter differences between the two such as the tip potential, local electric field, and hopping parameters, which are difficult to extract from the experiment.

We next obtain $g_v$ and demonstrate its gate tunability by performing linear fits to the split peaks in small $B$ for different $V_G$. The linear splitting is expected from the simple picture of coupling between $\vec{M}$ and $\vec{B}$. Figure 4c shows the zoom-in $dI^3/dV_S^3(V_S, B)$ around the valley split peaks at $V_G = 10$ V and 40 V, the nonlinear valley splitting is clearly visible. The yellow dashed lines are linear fits to the split peaks in small $B$ (additional data analysis details and data can be found in SI section S13). Based on the slopes of these fitted lines we extracted a $g_v = 1050 \pm 72$ at $V_G = 10$ V and a $g_v = 517 \pm 47$ at $V_G = 40$ V. This result demonstrates $g_v$ in ABA TLG is both giant and gate tunable, the combination of which is unparalleled in previously studied systems.[4, 16]

To compare the observed gate tunable $g_v$ with a theory based on a gapped MLG QD, we use plane wave representation $\psi(\vec{k})$ of the gapped MLG QD state at $B = 0$ T to estimate the valley g-factor as $g_v = \frac{2}{\mu_B} \int M(\vec{k}) |\psi(\vec{k})|^2 d\vec{k}$ (see SI section S7 for additional details). The calculated $g_v$ as a function of $\Delta$ for $\psi(\vec{r})$ with a Gaussian width of 150 nm and 300 nm are shown in Fig. 4d as a green solid line and blue solid line, respectively. We determined $\Delta$ at different $V_G$ by measuring the energy spacing between LL0- and LL0+ as shown in Fig. 4a. The experimentally extracted $g_v$ as a function of $\Delta$ are plotted in Fig.4d, the experiment and theory display good agreement. We notice the experimental $g_v$ at small $\Delta$ (i.e. small $V_G$) agrees better with theory that corresponds to larger Gaussian width for $\psi(\vec{r})$. This is consistent with the finding in Fig.2b that at lower $V_G$ the QD has weaker confinement.

In conclusion, we fabricated high quality ABA TLG/hBN heterostructure devices and studied their gate and sublattice resolved tunneling spectra in perpendicular electric and magnetic fields. Our work shows that the effective MLG bands of ABA TLG host giant and gate tunable topological magnetic moments that can generate large and tunable valley splitting in a small $B$. These findings demonstrate that ABA TLG is a unique platform for fabricating valley-based quantum information devices and studying topological magnetic moment related phenomena.

**Acknowledgments:** We acknowledge useful discussions with Angelika Knothe, Pablo Perez Piskunow, and Joseph Weston. J.V.J. and Z.G. acknowledges support from the National Science Foundation under award DMR-1753367. J.V.J acknowledges support from the Army Research Office under contract W911NF-17-1-0473. V.F. and S.S. acknowledge support from the European Graphene Flagship Core 3 Project. V.F. acknowledges support from Lloyd Register Foundation Nanotechnology Grant, EPSRC grants EP/V007033/1, EP/S030719/1 and EP/N010345/1. K.W. and T.T. acknowledge support from the Elemental Strategy Initiative conducted by the MEXT, Japan, Grant Number JPMXP0112101001, JSPS KAKENHI Grant Numbers JP20H00354.

**Figure 1**

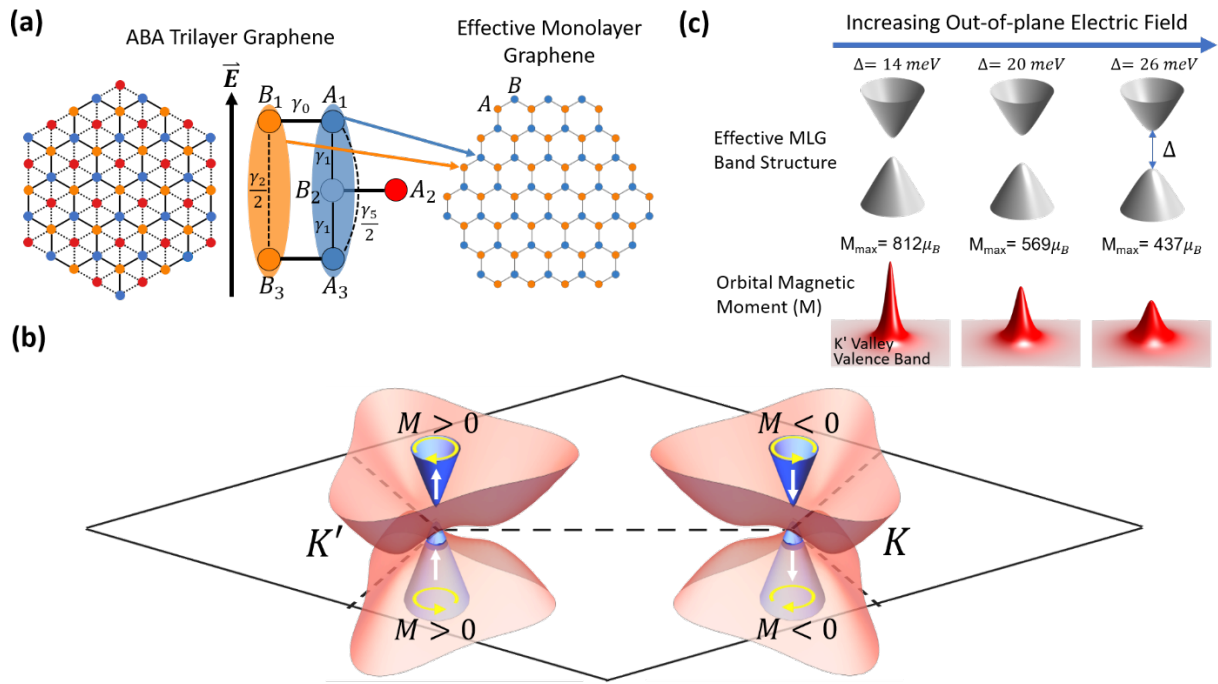

**Figure1: Effective MLG band in ABA TLG with giant and tunable topological magnetic moment**. **a,** Left panel: Top view of the ABA TLG atomic structure. Middle panel: Schematic of the ABA TLG unit cell and hopping parameters. Right panel: Mapping ABA TLG onto an effective MLG lattice. **b,** Schematic of the calculated low energy band structure of ABA TLG with no external electric field in K and K' valleys. Blue cones represent the effective MLG bands. The semi-transparent red shells represent the effective BLG bands. The yellow arrows depict the orientation of the self-rotating wave packet in each band and the white arrows correspond to the direction of the topological magnetic moment originating from the self-rotating wave packet. **c,** Upper panel: Low energy band structures of the effective gapped MLG with different out-of-plane electric fields applied to the ABA TLG. Lower panel: topological magnetic moment in the K' valley valence band of the corresponding gapped MLG bands shown in the upper panel. Here we assumed $v_F$ of the MLG band is $10^6\ m/s$.

**Figure 2**

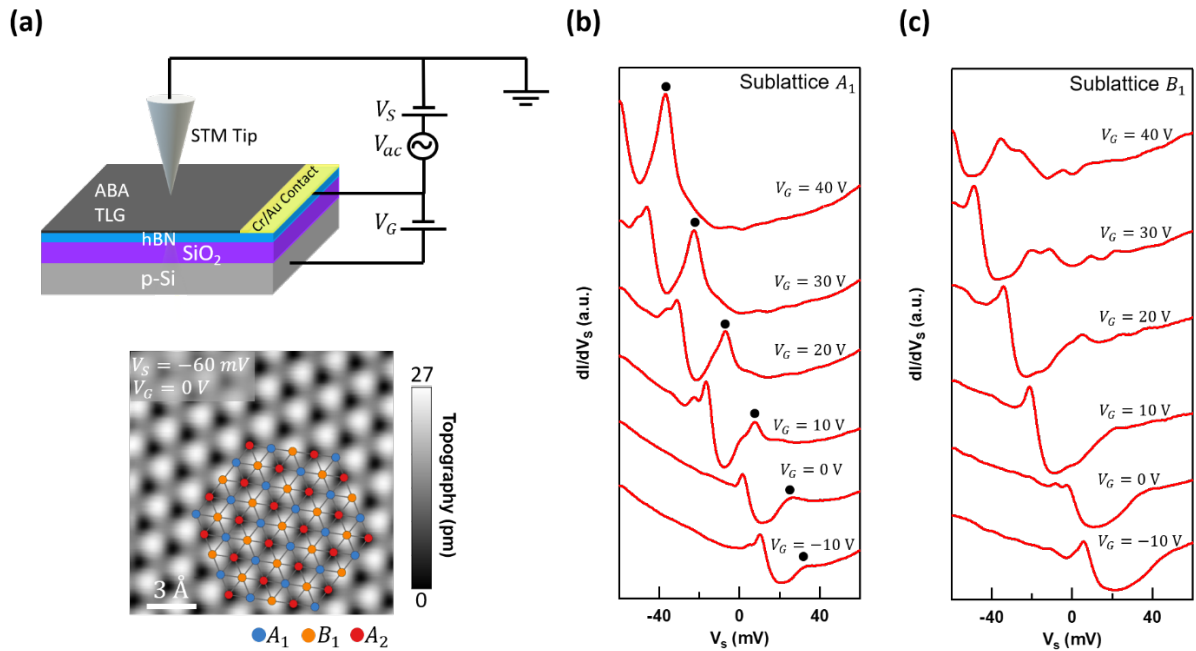

**Figure 2: Atomically resolved scanning tunneling spectroscopy (STS) of ABA TLG. a,** Upper panel: Schematic of the experimental setup. Lower panel: Atomically resolved topography of a pristine ABA TLG patch at $V_G = 0\,V$, the scanning parameters used are $I = 1$ nA, $V_S = -60$ mV. The ABA TLG atomic structure is overlaid on top of the topography, the definition of the sublattice is consistent with that in Fig. 1a and 1b. **b-c,** Tunneling spectra at various gate voltages on sublattice $A_1$ **(b)** and $B_1$ **(c)**. The set point used to acquire the tunneling spectra was $I = 1$ nA, $V_S = -60$ mV, with a 2 mV ac modulation.

Figure 3

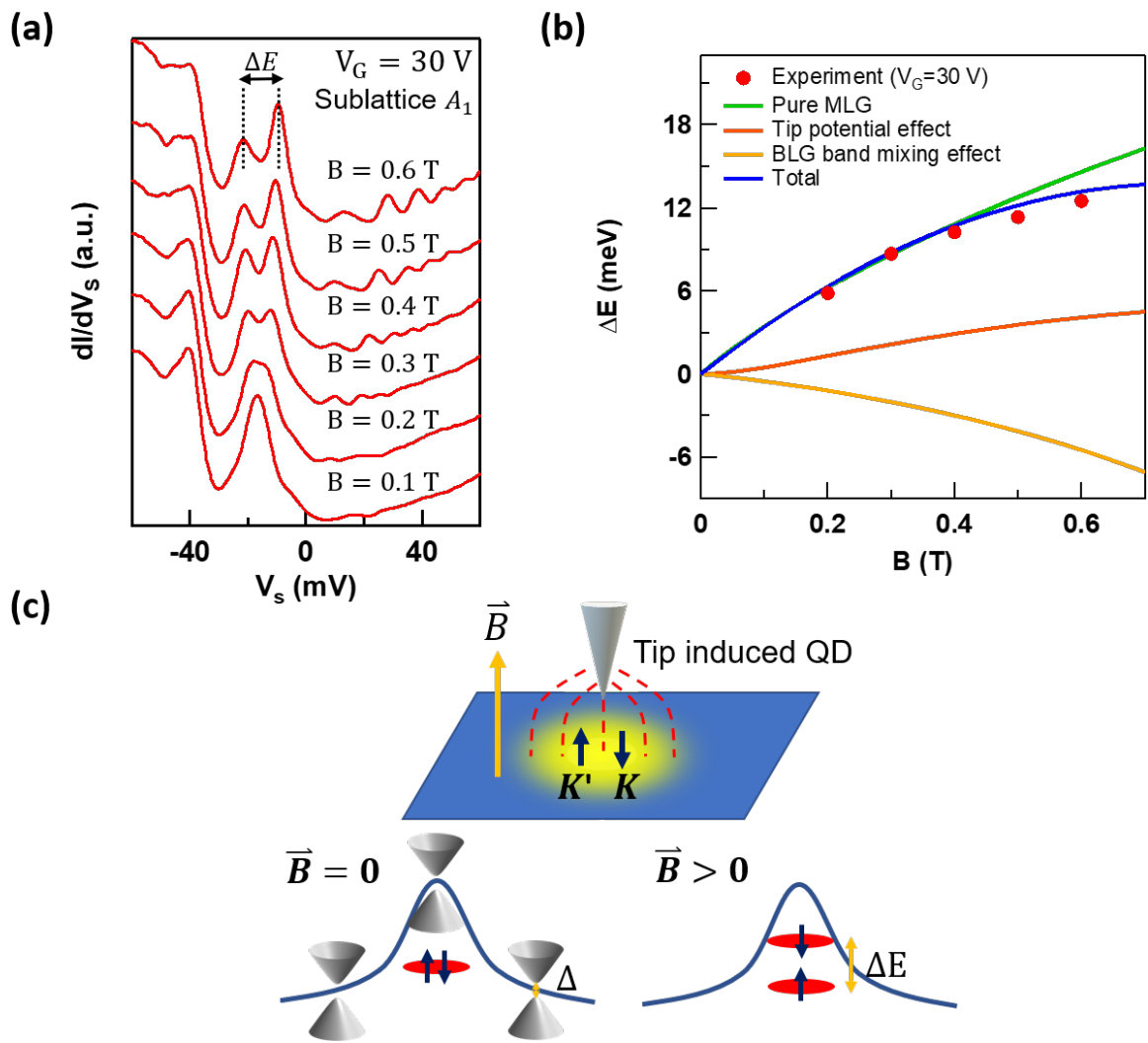

**Figure 3: Magnetic field controlled valley splitting in ABA TLG. a,** Tunneling spectra on sublattice $A_1$ at $V_G = 30\,V$ with different out of plane magnetic fields ($B$). The set point used to acquire the tunneling spectra was $I = 1\,\text{nA}$, $V_S = -60\,\text{mV}$, with a 2 mV ac modulation. **b,** Comparison between the experimental and theoretical valley splitting energy at $V_G = 30\,V$. The experimental splitting energy is extracted from **(a)**. The depth and width of the Gaussian potential well used in the theoretical calculation are 50 meV and 40 nm, respectively. **c,** Upper panel: Schematic of an STM tip induced QD in ABA TLG. The black arrows represent the directions of topological magnetic moments in TLG K and K' valleys, which can couple to an external out of plane *B* (orange arrow). Lower left panel: Schematic of the tip induced QD potential profile. The blue line represents the CNP of gapped MLG, the red oval schematizes the QD state arising from confinement. The black arrows represent the degenerate valley degree of freedom. Lower right panel: Schematic of QD state valley splitting under a *B*.

**Figure 4**

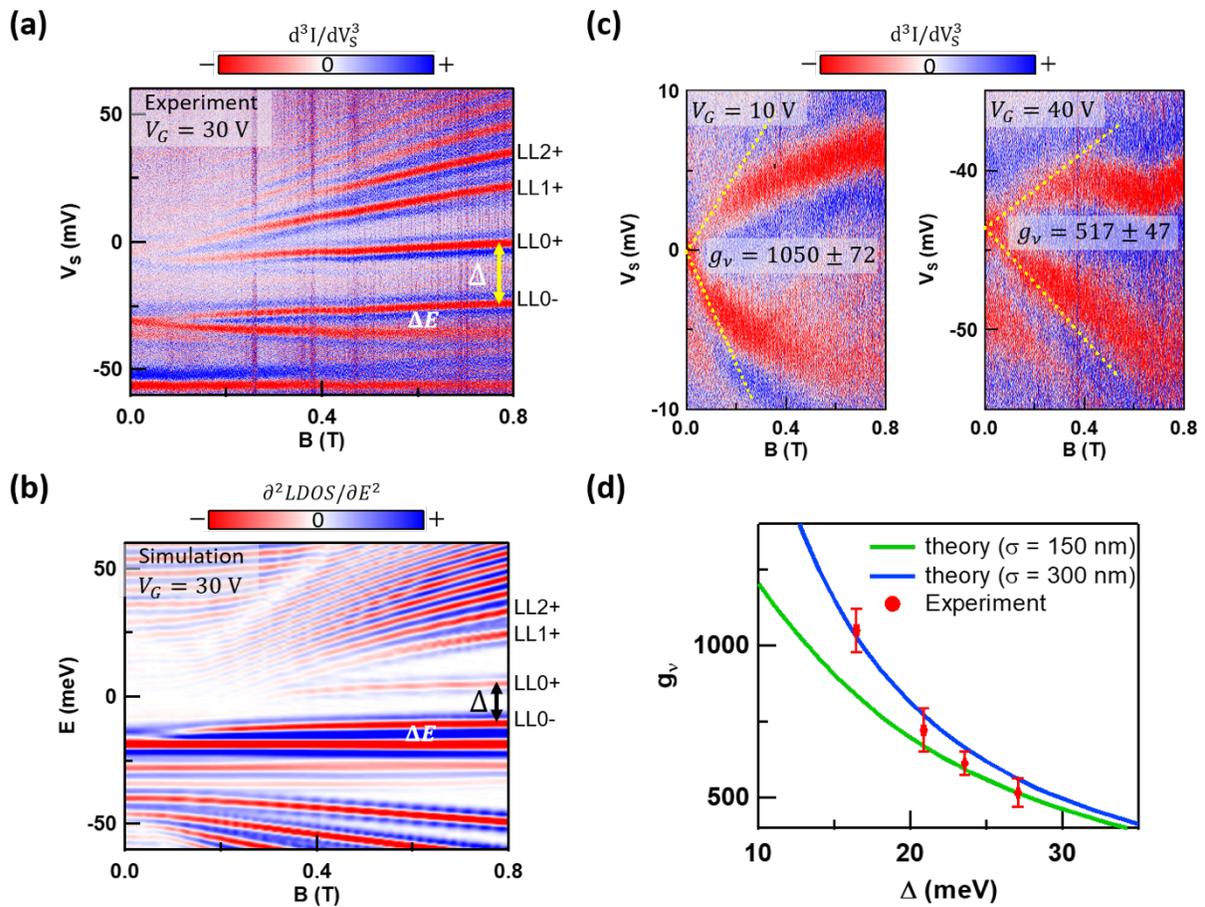

**Figure 4: Giant and gate tunable valley g factor in ABA TLG. a,** $d^3I/dV_S^3(V_S, B)$ at $V_G = 30\,V$, $d^3I/dV_S^3$ values were numerically calculated from measured $dI/dV_S$ values. The tunneling spectra were measured with a different calibrated STM tip and from a different location on the ABA TLG sample to where the data presented in Fig. 3 were acquired. The set point used to acquire the tunneling spectra was $I = 1$ nA, $V_S = -60$ mV, with a 2 mV ac modulation. The fan-like feature that appears at positive $V_S$ corresponds to MLG Landau levels. **b,** Simulated $\partial^2 LDOS(E, B)/\partial^2 E$ at $V_G = 30$ V. The depth and width of the Gaussian potential well used in the simulation are 30 meV and 100 nm, respectively and the $\gamma_2$ hopping was set to $-10$ meV. **c,** Zoom-in of the valley splitting peak in $d^3I/dV_S^3(V_S, B)$ at $V_G = 10$ V and $V_G = 40$ V. The yellow dashed lines are the linear fits to the splitting peaks near zero $B$. The data are acquired with the same STM tip and set point as in **a**. **d,** Experimental and theoretical $g_v$ as a function of $\Delta$ (MLG gap size). The experimental $\Delta$ value at different $V_G$ were extracted from the energy spacing between the LL0+ and LL0- in 0.8 T, which is schematized by the yellow arrow in **(a)**.

# Supporting Information

# Control of Giant Topological Magnetic Moment and Valley Splitting in Trilayer Graphene

Zhehao Ge[1,*], Sergey Slizovskiy[2,3,*], Frédéric Joucken[1], Eberth A. Quezada[1], Takashi Taniguchi[4], Kenji Watanabe[5], Vladimir I. Fal'ko[2,3,6,†], Jairo Velasco Jr.[1,†]

[1]Department of Physics, University of California, Santa Cruz, California, USA

[2]Department of Physics and Astronomy, University of Manchester, Oxford Road, Manchester, M13 9PL, UK

[3]National Graphene Institute, University of Manchester, Booth Street East, Manchester, M13 9PL UK

[4]International Center for Materials Nanoarchitectonics National Institute for Materials Science, 1-1 Namiki, Tsukuba, 305-0044, Japan

[5] Research Center for Functional Materials National Institute for Materials Science, 1-1 Namiki, Tsukuba, 305-0044, Japan

[6] Henry Royce Institute for Advanced Materials, Manchester, M13 9PL, UK

[*]These authors contributed equally to this manuscript.
[†]Email: jvelasc5@ucsc.edu, Vladimir.Falko@manchester.ac.uk

**Table of Contents**

**S1. Sample fabrication and STM measurements**

**Sample Fabrication**

ABA TLG was stacked on hBN using a standard polymer-based transfer method.[1] An ABA TLG flake exfoliated on a methyl methacrylate (MMA) substrate was mechanically placed on top of a $20-50$ nm thick hBN flake that rests on a $SiO_2/Si^{++}$ substrate where the oxide is 285 nm thick. Subsequent solvent baths dissolve the MMA scaffold. The TLG/hBN heterostructure is then annealed in forming gas (Ar/H$_2$) for six hours at 400 °C to reduce the amount of residual polymer left after the transfer of TLG. After that, an electrical contact to ABA TLG is made by thermally evaporating 7 nm of Cr and 200 nm of Au using a metallic stencil mask. The optical micrograph of the finished TLG/hBN device used in this work in shown in Fig. S1a. To further clean the sample's surface, the heterostructure is mechanically cleaned using an AFM tip.[2] Finally, the heterostructure is annealed under UHV at 400 °C for seven hours before being introduced into the STM chamber.

**STM Measurements**

The STM measurements were conducted in UHV with pressures better than $1 \times 10^{-10}$ mbar at 4.8 K in a Createc LT-STM. The bias is applied to the sample with respect to the tip. The frequency of the applied lock-in AC signal in the circuit is 704 Hz. The STM tips were electrochemically etched tungsten tips and calibrated on a clean Au(111) surface to ensure that the tip was free from artifacts.[3,4] All topography map images were plotted by WSxM software.[5]

Before the gate and magnetic field resolved STS measurements, the surface cleanness of a 20 nm × 20 nm window is always checked. A typical topography of a clean 20 nm × 20 nm TLG region is shown in Fig. S1b.

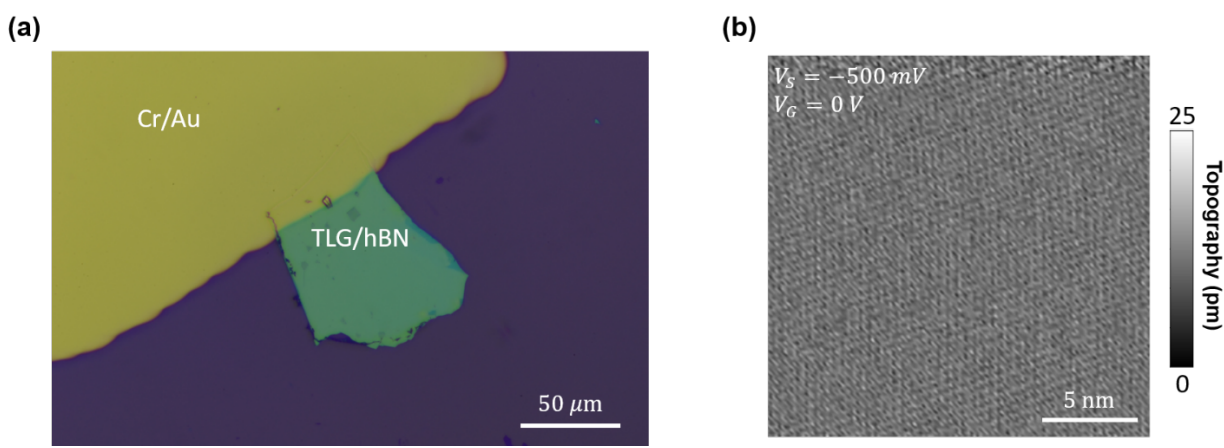

**Figure S1: Optical micrograph of the ABA TLG device and STM topography of a pristine region on the device. a,** Optical micrograph of the ABA TLG device used for gate and $B$ resolved STS study in this work. TLG covered the entire hBN substrate and is connected to a Cr/Au contact. **b,** STM topography of a 20 x 20 nm pristine ABA TLG area on the device acquired at $V_G = 0$ V. The scanning parameters used are $I = 10$ pA, $V_S = -500$ mV.

## S2. Large scale atomically resolved topography

To rule out the presence of moiré patterns with periodicity larger than the scanning window size in Fig. 2a, atomically resolved topography were acquired. Figure S2 shows a $10 \times 10$ nm² atomically resolved topography for ABA TLG. No moiré pattern was observed. Thus, we conclude the ABA TLG is misaligned with the hBN substrate.

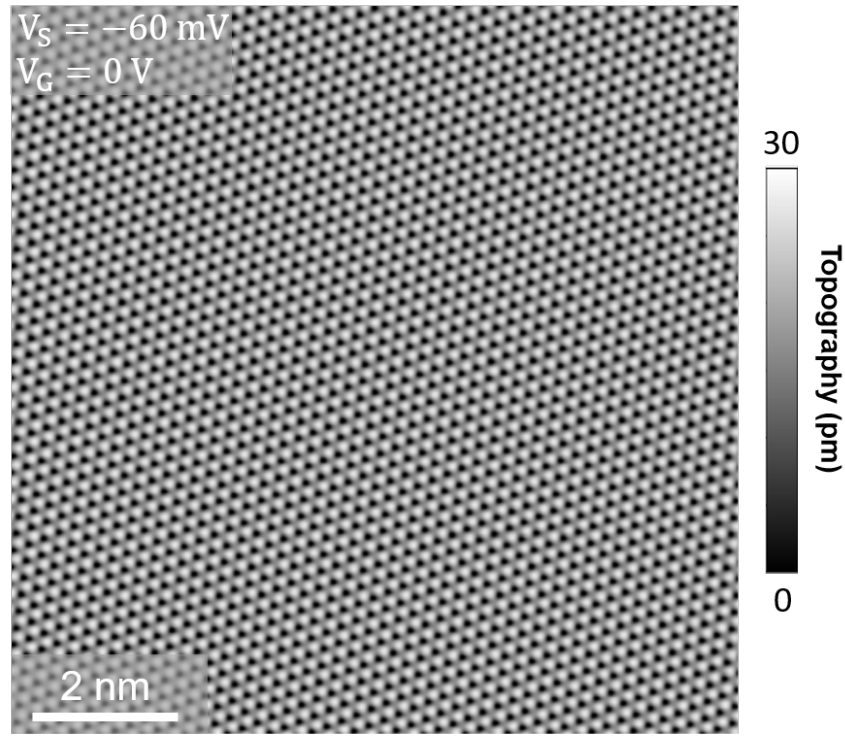

**Figure S2: $10 \times 10$ nm² atomically resolved STM topography.** Atomically resolved topography of a pristine ABA TLG area at $V_G = 0\ V$, the scanning parameters used are $I = 1\ nA$, $V_S = -60\ mV$.

## S3. Sublattice identification

The topography in Fig. 2a is acquired through the constant tunneling current mode of the STM. According to the Tersoff-Hamann theory[6] the tunneling current at $B = 0$ T is

$$I \propto e^{-\alpha z(r)} \int_{E_F}^{E_F+eV_S} LDOS(E,r)dE.$$

In the constant tunneling current mode the above expression is equal to a constant ($I_0$). Therefore, the tip sample distance at each location can be written as:

$$z(r) \propto \frac{1}{\alpha}[\ln\left(\int_{E_F}^{E_F+eV_S} LDOS(E,r)dE\right) - \ln(I_0)].$$

As a result, a location with larger $\int_{E_F}^{E_F+eV_S} LDOS(E,r)dE$ value will have a larger z value in an STM topography.

In Fig. S3, we show the calculated LDOS of sublattice $A_1$ and $B_1$ near the TLG charge neutrality point. Because the LDOS between the Fermi level and applied bias ($V_S = -60mV$) at sublattice $B_1$ is always larger than the sublattice $A_1$ LDOS, the integral of the LDOS between the Fermi level and applied bias for sublattice $B_1$ will be larger than that of sublattice $A_1$. As a result, the atom with the highest intensity in Fig. 2a corresponds to sublattice $B_1$ and the atom with lower intensity corresponds to sublattice $A_1$. In addition, since the tunneling current is mainly due to contributions from the top layer of TLG, the dark spot in Fig. 2a corresponds to the position of sublattice $A_2$, which lies in the middle layer.

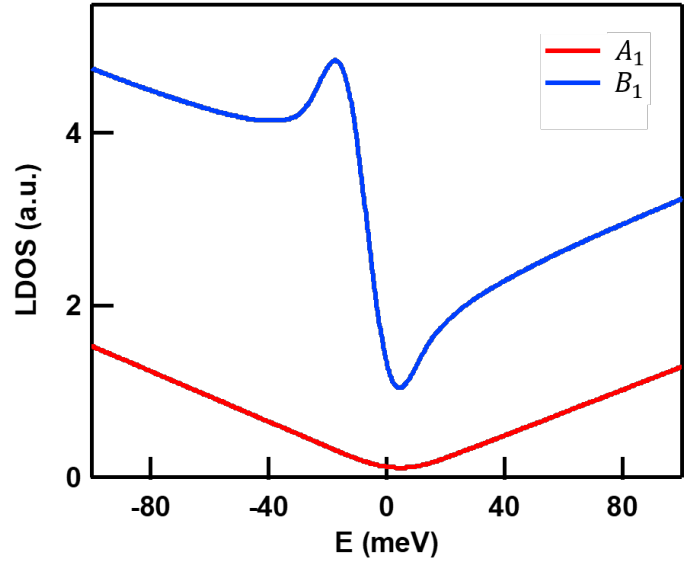

**Figure S3: Calculated tight-binding local density of states on sublattice $A_1$ and $B_1$ of pristine ABA TLG.** The hopping parameters and onsite energies used in the tight-binding model are the same as described in the supporting information section 7. Sublattice $B_1$ has greater intensity than sublattice $A_1$.

### S4. STS data of Fig.2b and 2c before averaging

Figure S4 shows raw selected STS data before an averaging of these spectra was performed. These spectra were measured at different locations that correspond to sublattice $A_1$ and sublattice $B_1$. These measurements were taken at different $V_G$ with $B = 0$ T. The $dI/dV_S$ peaks that are related to our main findings exist in the data before averaging and agree well with the data after averaging, which are shown in Figs. 2b and 2c. These results demonstrate the high reproducibility of the tunneling spectra at different $V_G$ and different sublattices.

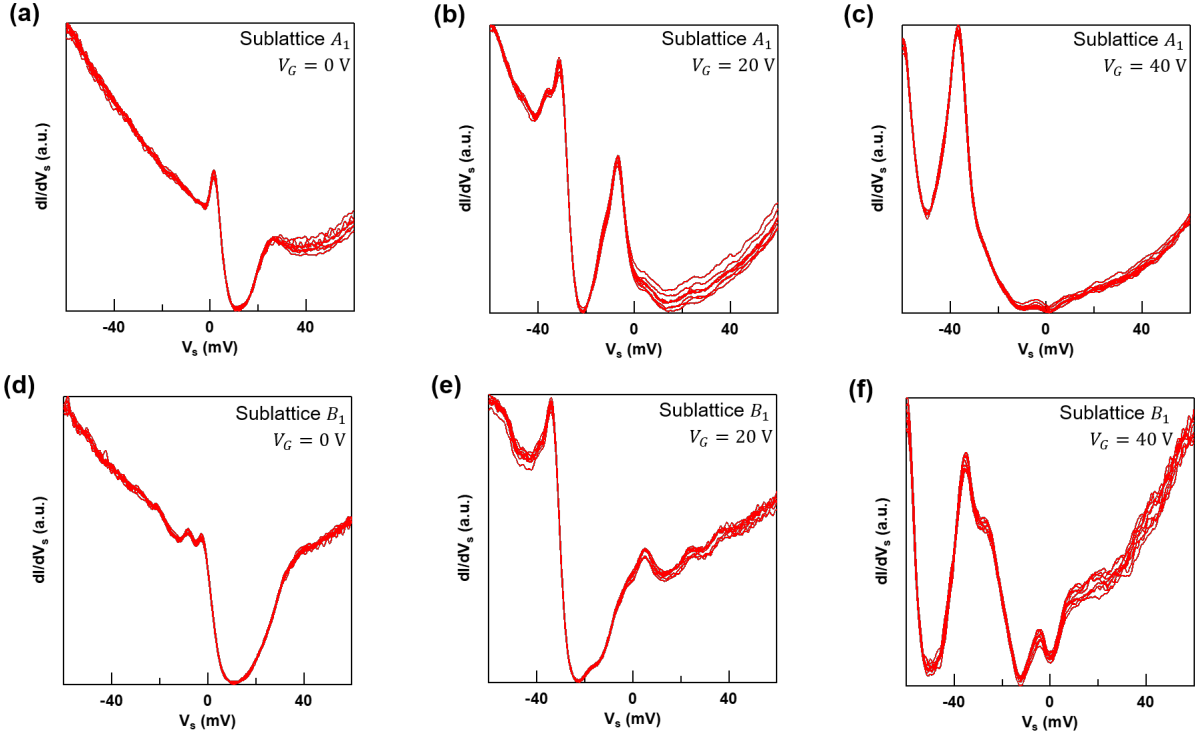

**Figure S4: Reproducibility of sublattice resolved tunneling spectra. a-c,** Tunneling spectra measured from sublattice $A_1$ distributed at nine different locations within a $1.6 \times 1.6$ nm² window at $V_G = 0$ V, $V_G = 20$ V and $V_G = 40$ V, respectively. **d-f,** Tunneling spectra measured from sublattice $B_1$ distributed at nine different locations within a $1.6 \times 1.6$ nm² window at $V_G = 0$ V, $V_G = 20$ V and $V_G = 40$ V, respectively. The set point used to acquire the tunneling spectra in a-f was $I = 1$ nA, $V_S = -60$ mV, with a $2mV$ ac modulation.

## S5. $B$ field dependent STS data on sublattice $A_1$ and $B_1$ at various $V_G$

Figure S5 shows the experimentally measured tunneling spectra on sublattice $A_1$ and $B_1$ in different $B$ and at different $V_G$. A peak splitting is observed on the tunneling spectra of sublattice $A_1$ at different $V_G$ in finite $B$, whereas this peak splitting phenomenon is not observed on sublattice $B_1$. Additional satellite $dI/dV_S$ peaks are visible and correspond to MLG Landau levels. These peaks are observed on both sublattice $A_1$ and $B_1$ in finite $B$ at different $V_G$.

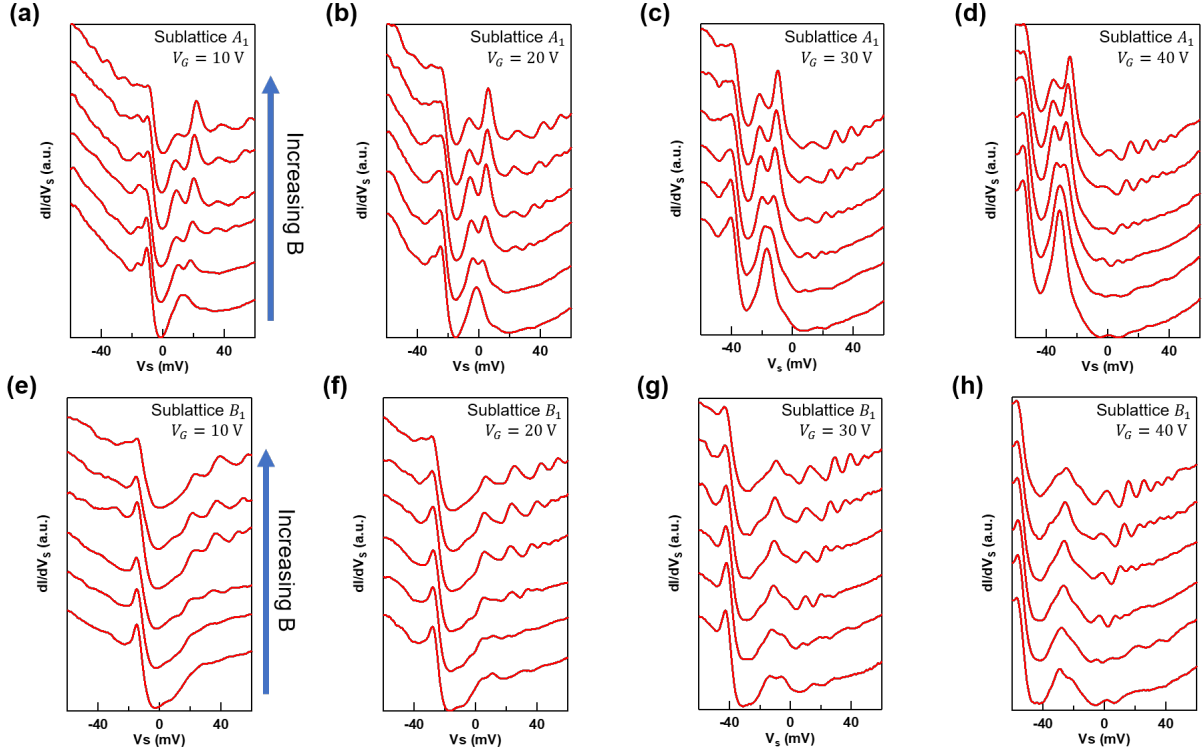

**Figure S5:** *B* **dependent tunneling spectra on sublattice $A_1$ and $B_1$ at different $V_G$. a-d,** Tunneling spectra measured from sublattice $A_1$ at $V_G = 10$ V, $V_G = 20$ V, $V_G = 30$ V and $V_G = 40$ V, respectively. Within each panel and from bottom to top, the applied *B* increases from 0.1 T to 0.6 T with a 0.1 T interval. **e-h,** Tunneling spectra measured from sublattice $B_1$ at $V_G = 10$ V, $V_G = 20$ V, $V_G = 30$ V and $V_G = 40$ V, respectively. Within each panel and from bottom to top, the applied *B* increases from 0.1 T to 0.6 T with a 0.1 T interval. The set point used to acquire the tunneling spectra in **a-h** was $I = 1\ nA$, $V_S = -60\ mV$, with a $2mV$ ac modulation.

## S6. Band gap extraction of the effective MLG bands at different $V_G$

At high *B*, the effect of magnetic field confinement is more significant than the QD confinement. As a result, Landau levels dominate the tunneling spectra in this regime. In gapped MLG, a zeroth LL (LL0) exists near the band edges of the conduction (labelled LL0+) and valence (labelled LL0-) bands, note that LL0+ and LL0- are in the different valleys. So, we use the energy spacing of LL0+ and LL0- to estimate the band gap size of the effective MLG bands in ABA TLG.

Figure S4a shows an experimentally measured $dI/dV_S(V_S, V_G)$ plot in $B = 1\,\text{T}$, LLs from the effective MLG bands can be clearly observed. The LLs in the conduction band are labeled up to the 5th LL.

We next extract the dispersion of the LLs as a function of $B$. To do this we acquire a vertical line cut from $dI/dV_S(V_S, V_G)$ measurements in different $B$. Figure S6b shows the dispersion of different LLs at $V_G = 40\,V$ and with increasing $B$, notably the horizontal axis is plotted as $\sqrt{B}$. Evidently, the energy dispersions of LL0+ and LL0- are nearly $B$ independent. In contrast, the dispersions of the other five LLs are approximately linearly dependent on $\sqrt{B}$ at high $B$. This behavior agrees with the expected LL energy dispersion in MLG.[7] Figure S4c shows the extracted band gap size at selected $V_G$. As can be seen from Fig. S6a, the band gap (depicted as doubled sided yellow arrows) increases with increasing $V_G$. Figure S6c reveals that the band gap size roughly has a linear dependence on gate voltage. Representative tunneling spectra that were used to extract the gap values are shown in Fig. S7, where the LL0+ and LL0- are indicated by black arrows. These representative spectra were acquired at several different $V_G$ and with $B = 1\,\text{T}$.

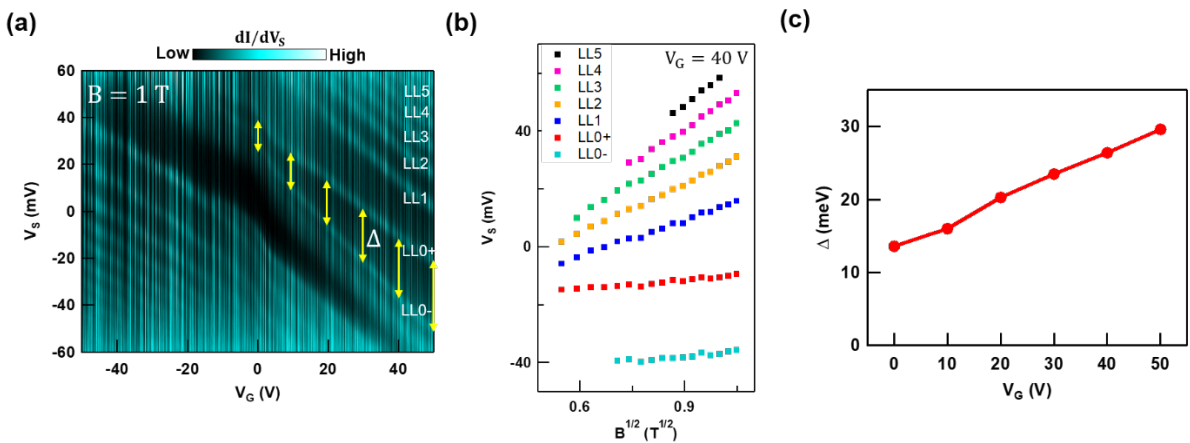

**Figure S6: Effective MLG band gap extraction. a,** Measured $dI/dV_S(V_S, V_G)$ in $B = 1\,\text{T}$ at $V_G = 40$ V. The tunneling spectra were measured with a different calibrated STM tip and acquired from a different location on the ABA TLG sample compared to the data presented in Fig. 3 and Fig. 4. The tunneling spectra measured by this STM tip did not have strong sublattice dependence. Nonetheless, trends consistent with experimental signatures discussed in the main text were seen

with this tip. The set point used to acquire the tunneling spectra was $I = 1$ nA, $V_S = -60$ mV, with a 2 mV ac modulation. The yellow double-sided arrows represent the gap size of the effective MLG bands at different $V_G$. **b,** Extracted Landau level dispersion at $V_G = 40$ V from $dI/dV_S(V_S, V_G)$ in different $B$. The extracted LL dispersions are plotted as a function of $\sqrt{B}$. **c,** Extracted gap size of the effective MLG bands at different $V_G$ from **a**.

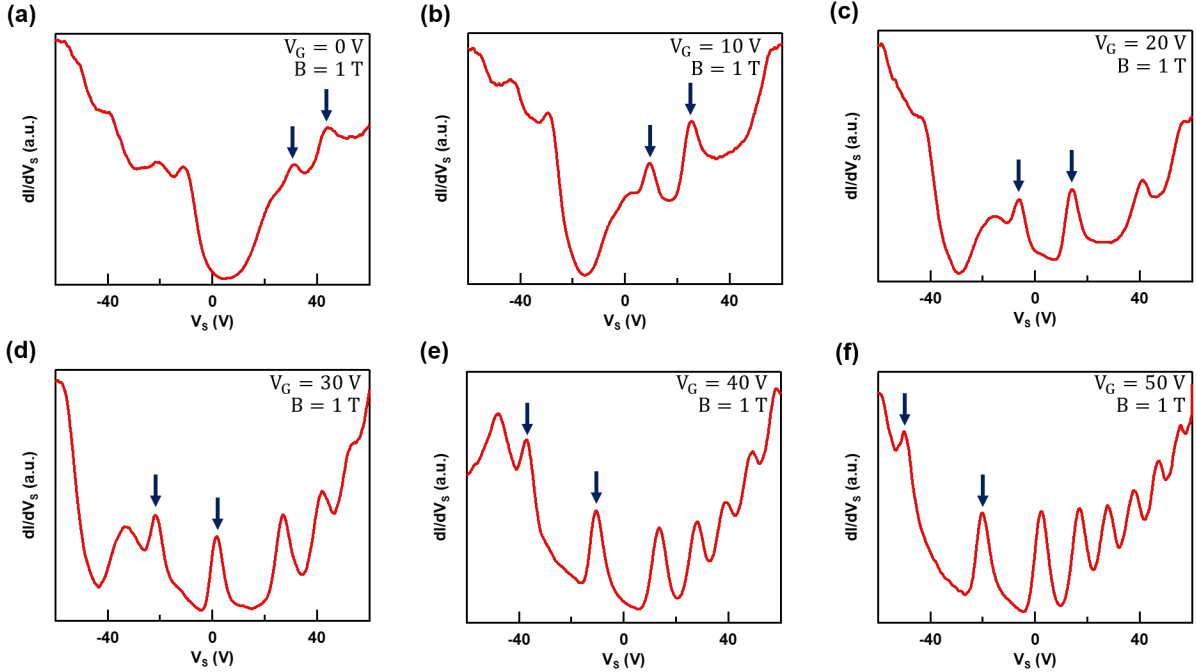

**Figure S7: Tunneling spectra in $B = 1$ T at different $V_G$. a-f,** Tunneling spectra extracted from Fig. S4a at $V_G = 0$ V, 10 V, 20 V, 30 V, 40 V and 50 V, respectively. The LL0+ and LL0- are indicated by black arrows. To reduce noise, each spectra consist of an average of spectra within a gate voltage range of ±0.4 V near the targeted $V_G$. The set point used to acquire the tunneling spectra was $I = 1$ nA, $V_S = -60$ mV, with a 2 mV ac modulation.

## S7. Theoretical model of STS on ABA TLG

The effect of STS tip and back-gate can be modeled by introducing electrostatic potentials for electrons on the layers of ABA TLG, $U_1, U_2, U_3$, where the index 1 (3) refers to the top (bottom) layers. The potentials computed via self-consistent Hartree screening procedure, taking into account both the redistribution of charge between graphene layers and the polarizability of carbon orbitals in graphene[8] can be linearized (for $U_i \ll \gamma_1$) as $\delta U_2 = 0.7 \delta U_3$ and $\delta U_1 = 0.6 \delta U_3$: this takes into account the influence of the bottom gate and, similarly, $\delta U_2 = 0.7 \delta U_1$ aand $\delta U_3 = 0.6 \delta U_1$ for the effect of the tip potential. In addition, we account for a contribution $U_2^{(0)} \approx -5$ meV to $U_2$ (determined by fitting to include energy difference between the middle and outer (top/bottom) graphene layers. Calculating the total electron density allows us to identify $U_3 \cong -0.0019 \, V_G$.

Below we use the above parameter and the Tight Binding (TB) model of ABA TLG to describe the structure of low-energy bands of TLG, identifying effective "gapped monolayer" and "bilayer" type subbands and explaining the origin of experimentally observed features.

**monolayer- and bilayer-type subbands**

Below, we use the following notation: the sublattices of ABA TLG are recorded as $A_1$, $B_2$, $A_3$ for the trimer chain of carbons located one under another, and as $B_1$, $A_2$, $B_3$ for non-trimer carbons, located above/below the centers of honeycombs, see Fig. S8. We use the standard Slonczewski-Weiss-McClure tight-binding model to describe graphene multilayers.[9-12] The nearest neighbor $A_i$ to $B_i$ in-plane hopping $\gamma_0 \approx 3.16$ eV determines graphene Fermi velocity. The parameter $\gamma_1 \approx 0.39$ eV denotes vertical hopping along trimers, direct $A_1$ to $A_3$ hopping is $\gamma_5/2$ ($\gamma_5 \approx 0.044$ eV in graphite), direct $B_1$ to $B_3$ hopping is $\gamma_2/2$ (we choose $\gamma_2 \approx -0.02$ eV if not specifically stated), skew next-layer trimer to non-trimer hopping $\gamma_4 \approx 0.14$ eV is responsible for a small electron-hole asymmetry and skew next-layer non-trimer to non-trimer hopping $\gamma_3 \approx -0.38$ eV leads to trigonal warping of bands. Note that energy difference between orbitals of trimer and non-trimer carbons, $\Delta_{AB} \approx 0.025$ eV turns out to be important for the discussion below.

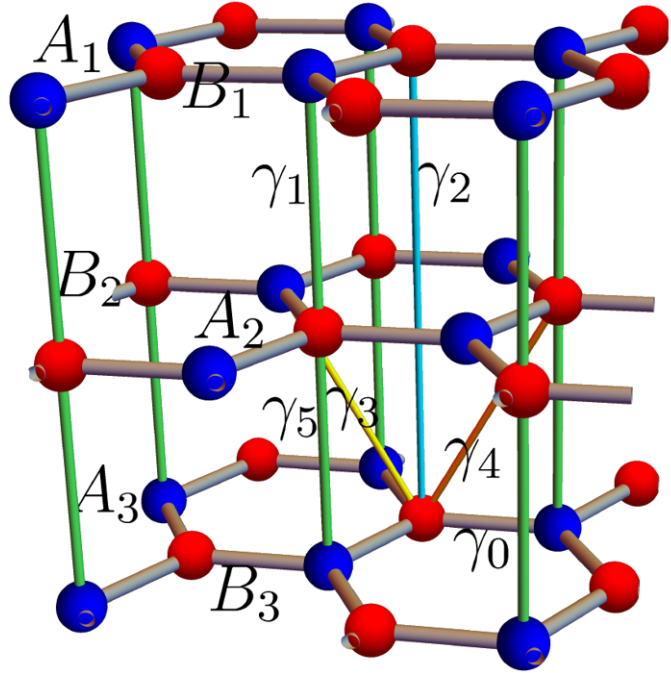

**Figure. S8: Structure and hopping integrals of Bernal stacked trilayer graphene.**

Considering the vertical hopping in the trimer chains, $\gamma_1$, as a large energy scale, we note that there is one low-energy eigenstate of a trimer chain: $A_1 - A_3$ with energy determined by poorly-known $A_1 - A_3$ hopping integral $\gamma_5/2$ (we denote by $A_i$ and $B_i$ the corresponding $\pi_z$ orbitals considered as basis states in a tight-binding model). At the $K/K'$ points

$$E_{A_1-A_3}(k=0) = \frac{U_1(x)+U_3}{2} - \frac{1}{2}\gamma_5 + \Delta_{AB} \tag{1}$$

The other low-energy orbitals at $k = 0$ are $A_2$-based orbitals and a linear combination of $B_1$ and $B_3$ orbitals,

$$B_{13\pm} \sim \left(U_1 - U_3 \mp \sqrt{\gamma_2^2 + (U_1 - U_3)^2}\right) B_1 + \gamma_2 B_3 \tag{2}$$

with energies,

$$E_{B_{13\pm}}(k=0) = \frac{U_1(x)+U_3}{2} \mp \frac{1}{2}\sqrt{\gamma_2^2 + (U_1 - U_3)^2}. \tag{3}$$

Noting that $\gamma_2 < 0$, we find that for $|U_1 - U_3| \leq |\gamma_2|$, the combination $B_{13-}$ tends to an antisymmetric combination $B_1 - B_3$ and thus the in-plane hopping couples it to $A_1 - A_3$. So, $A_1 - A_3$ and $B_{13-}$ form the two sublattices of an effective gapped graphene monolayer subsystem. The matrix element of in-plane hopping between $E_{A_1-A_3}$ and $E_{B_{13-}}$ orbitals determines an effective Dirac velocity,

$$(v_{\text{Dirac}})^2 = v^2 \left( \frac{1}{2} + \frac{1}{2} \frac{-\gamma_2}{\sqrt{(U_1-U_3)^2+\gamma_2^2}} \right) \approx v^2 \left( 1 - \frac{(U_1-U_3)^2}{4\gamma_2^2} \right) \quad ,$$

where $v \approx 10^6$ m/s is a Fermi velocity of graphene monolayer. The relative position of $E_{A_1-A_3}$ and $E_{B_{13-}}$ at the K point determines the gap size of an effective Dirac dispersion embedded into TLG,

$$\Delta = E_{B_{13-}}(k=0) - E_{A_1-A_3}(k=0) = \frac{1}{2}\sqrt{\gamma_2^2 + (U_1-U_3)^2} + \frac{1}{2}\gamma_5 - \Delta_{AB} \quad . \quad (4)$$

Combining the above equations, we arrive at,

$$(v_{Dirac})^2 = \frac{v^2}{2}\left(1 + \frac{-\gamma_2}{2\Delta+2\Delta_{AB}-\gamma_5}\right) \quad , \quad (5)$$

which is useful since it eliminates the unknown magnitude of the tip potential in favor of directly measurable $\Delta$. Using the relation $U_1 - U_3 \cong 0.4(\delta U_{1tip} + 0.0019\, V_G)$, we find that the gap between the monolayer dispersion branches grows with the gate voltage. Assuming $\gamma_5 = 44$ meV, as in graphite, we need to choose $\Delta_{AB} \approx 21$ meV to reasonably account for the experimental results. Note that the typical values for sublattice energy difference found in the literature for the bulk graphite are $\Delta_{AB} \approx 60$ meV, which would lead to $\Delta < 0$ and contradicts to the presented experimental results, however, it is only $\Delta_{AB}$ for the surface layers that really matters in our calculation, and at the surface the $A$ sites have vertical hybridization only in one direction, justifying smaller value for $\Delta_{AB}$.

The effective bilayer branches are localized on $B_1$, $B_3$ and $A_2$ sublattices. The size and the sign of the gap between the above bilayer branches is determined by the gate potential and the localized tip potential

$$E_{B_{13+}} - E_{A_2} = \frac{U_1+U_3}{2} - \frac{1}{2}\sqrt{\gamma_2^2 + (U_1 - U_3)^2} - U_2 - \Delta_{AB} \tag{6}$$

When a back-gate voltage breaks the inversion symmetry, the BLG dispersion branch localized on $B_1$, $B_3$ hybridises with Dirac dispersion branch, while the $A_2$ branch remains intact, see Fig. S13. This hybridization of bilayer and monolayer bands is specified by the matrix element

$$\langle A_1 - A_3|H|B_{13+}\rangle = \hbar\sqrt{v^2 - v_{Dirac}^2}\,(k_x + i\,k_y) \tag{7}$$

which, according to the second order perturbation theory, leads to the correction

$$\delta E(k) = \frac{\hbar^2 k^2 (v^2 - v_{Dirac}^2)}{\Delta + 2\Delta_{AB} - \gamma_5 - \hbar k\, v_{Dirac}} \tag{8}$$

to the energy of the valence band of the gapped monolayer dispersion.

**Gapped Dirac dispersion in the tip potential**

Prior to presenting calculations for the full TLG model, let us discuss the expected STS spectrum of an effective gapped Dirac system (effective gapped graphene monolayer) in a circularly-symmetric potential. We assume that the STS tip is probing the local density of states in the center of a polar-symmetric potential, created by the tip. In polar coordinates, the Dirac equation in magnetic field reads[13,14]:

$$\begin{pmatrix} \xi\Delta/2 & -\hbar v_F e^{-i\phi}(i\partial_r + \frac{\partial_\phi}{r} + \frac{ir}{2l_B^2}) \\ \hbar v_F e^{i\phi}(-i\partial_r + \frac{\partial_\phi}{r} + \frac{ir}{2l_B^2}) & -\xi\Delta/2 \end{pmatrix}\Psi = (E - U(r))\Psi \tag{9}$$

where $l_B = \sqrt{\hbar/(eB)} \approx 26/\sqrt{B/(\text{Tesla})}$ nm and $\xi = \pm 1$ refers to $K$ and $K'$ valleys, $v_F$ is a Fermi velocity of ungapped Dirac cone and $\Delta$ is a bandgap. It is convenient to expand the wave-function on each of the sublattices in the basis of

$$|n,m\rangle = \psi_{n,m}(r,\phi) = e^{im\phi} g_{n-\frac{m+|m|}{2},m}(r/l_B) \tag{10}$$

for $n \geq 0$ and $m < n$ and where an oscillator radial eigenfunction g is defined as

$$g_{n,m}(r) = e^{-\frac{r^2}{4}} r^{|m|} \sqrt{\frac{2^{-|m|}(|m|+n)!}{2\pi n!(|m|!)^2}} F_1\left(-n; |m|+1; \frac{r^2}{2}\right)$$

The off-diagonal matrix elements act as raising and lowering operators in this basis and change the angular momentum $m$ by one:

$$-\hbar v_F e^{-i\phi}\left(i\partial_r + \frac{\partial_\phi}{r} + \frac{ir}{2l_B^2}\right)|n,m\rangle = -i\hbar v_F \sqrt{2n}|n-1,m-1\rangle \tag{11}$$

so, we can define a lowering operator $a = -\frac{i}{\sqrt{2}}\left(i\partial_r + \frac{\partial_\phi}{r} + \frac{ir}{2l_B^2}\right)$ so that

$a|n,m\rangle = \sqrt{n}|n-1,m-1\rangle$ and rewrite the Hamiltonian as

$H = \begin{pmatrix} \xi\Delta/2 & -i\sqrt{2}\hbar v_F a \\ i\sqrt{2}\hbar v_F a^\dagger & -\xi\Delta/2 \end{pmatrix}$. Only the $m=0$ elements of the basis set have non-zero value at the origin, $r=0$, where we want to compute the LDOS.

In the absence of any tip potential, the solutions of Eq. (11) are:

$$\Psi_{0,m,\xi} = \begin{pmatrix} 0 \\ |n,m\rangle \end{pmatrix} \quad , \quad E_{0,m,\xi} = -\xi\Delta/2$$

(12)

$$\Psi^\pm_{n>0,m,\xi} \sim \begin{pmatrix} i\sqrt{2n}\hbar v_F/l_B |n-1,m-1\rangle \\ (\xi\Delta/2 - E^\pm_{n,m,\xi})|n,m\rangle \end{pmatrix}, \quad E^\pm_{n>0,m,\xi} = \pm\sqrt{\left(\frac{\Delta}{2}\right)^2 + 2n\left(\frac{\hbar v_F}{l_B}\right)^2} \tag{13}$$

A smooth, radially symmetric, positive tip potential introduces matrix elements between basis elements with different $n$ but having the same $m$, so, it is sufficient to restrict attention to $m=0$ and $m=1$ basis states to get LDOS at $r=0$. In the leading order of perturbation theory, the strongest matrix element occurs for the $|0,0\rangle$ basis element, which corresponds to wave-function having the largest overlap with the tip potential. Higher orders of perturbation theory non-linearly enhance the effect of this matrix element for the valence band states, because the further the tip potential drags the valence band states into the gap, the more localized the corresponding wave-function becomes near the tip. In the valence band, the $|0,0\rangle$ state is present in $\Psi_{0,0,K+}$ and $\Psi^-_{1,1,K-}$ on the B sublattice. These two levels are the first to be dragged into the gap and they are degenerate at $B = 0$, corresponding to valley-degeneracy expected in the absence of magnetic field. The relevant A sublattice based state is $\Psi^-_{1,0,K-}$ (the $|0,0\rangle$ wave function in the A sublattice is shifted into the gap by the tip potential, Fig. S9).

At a non-zero B, levels $\Psi_{0,0,K+}$ and $\Psi^-_{1,1,K-}$ split, with the splitting given by

$$E_{0,0,K+} - E^-_{1,1,K-} = \sqrt{\left(\frac{\Delta}{2}\right)^2 + 2\left(\frac{\hbar v_F}{l_B}\right)^2} - \frac{\Delta}{2} \approx \frac{2\hbar e v_F^2}{\Delta} B,$$

(14)

where we neglect the effect of the tip potential. This estimation agrees with the splitting that would arise from the valley-contrasting topological magnetic moment for gapped MLG, as given in[15],

$$m(k) = \pm \frac{e}{\hbar} \frac{\Delta}{[\Delta/(\hbar v_{Dirac})]^2 + 4k^2}, \quad (15)$$

and evaluated at $k = 0$, with $v_F = v_{Dirac}$ in Eq. (4). Taking tip potential into account, we note that the wave-functions localized by the tip potential are not plane waves even at $B = 0$. We assume a Gaussian shape of the wave-functions broadened to $\sigma \sim 50$ nm (which is of the order of the tip potential radius). Averaging topological magnetic moment in Eq.(15) over the wave-function, we get for the g-factor:

$$g = 2\mu_B^{-1} \int d^2k \, m(k) \frac{\sigma^2}{2\pi} e^{-\sigma^2 k^2/2} = \frac{-\Delta \sigma^2 \exp(f) Ei(f)}{4\hbar \mu_B}, \text{ where } f = \frac{\Delta^2 \sigma^2}{8\hbar v_{Dirac}^2} \quad (16)$$

that is plotted in Fig. 4d alongside the experimental results.

Finally, we note that the level-splitting $E_{0,0,K+} - E^-_{1,1,K-}$ becomes non-linear with growing magnetic field: the nonlinearity in eq.(14) becomes essential when the splitting becomes comparable to $\Delta/2$. There are other contributions to the level-splitting that also contribute to non-linearity: One comes from mixing with bilayer band, that decreases the level splitting as described by eq.(8): in the finite magnetic field this leads to a correction $\delta E\left(\frac{1}{l_B\sqrt{2}}\right)$ to the energy $E^-_{1,1,K-}$ of -1st Landau level, leading to reduction of splitting at larger magnetic fields. Another contribution appears due to a positive STS tip potential. According to eqs. (12) and (13), the levels $\Psi_{0,0,K+}$ and $\Psi^-_{1,1,K-}$ have the same $|0,0\rangle$ wave-function (albeit in different valleys) at $B = 0$, but as $B$ grows, the state $\Psi^-_{1,1,K-}$ is getting an increasing $|1,1\rangle$ component of wave-function that has broader spatial support. As a result, at $B > 0$, the influence of the tip potential is weaker for the level —in comparison to $\Psi^-_{0,0,K+}$. This effect can be computed in the leading order of the perturbation theory (we assumed a Gaussian potential shape). This results in a positive contribution to the splitting, which is most pronounced when the magnetic length is of the order of the tip potential radius. The three contributions are plotted and compared in Fig.3b of the main text.

**Scaled tight-binding model**

To solve the STS problem in the full generality, considering the full dispersion of TLG, we used the scaled tight-binding model, where the low-energy spectrum near $K$ and $K'$ points is modeled with the help of an effective tight-binding model on TLG lattice with increased (up-scaled) lateral distances. We note that the low-energy dispersion obtained in $k \cdot p$ expansion in the vicinity of $K, K'$ points is written in terms of velocities $v_i = \frac{\sqrt{3}}{2} a \gamma_i, i = 0,3,4$, vertical hoppings $\gamma_{1,2,5}$ and sublattice energy difference $\Delta_{AB}$. So, one can increase the lattice spacing ($a \approx 0.246$ nm) as $a \to as$ and simultaneously decrease the hoppings $\gamma_i \to \frac{\gamma_i}{s}, i = 0,3,4$ to preserve the low-energy dispersion. We used a scaling factor $s = 10$ in our calculations performed with the help of Kwant package[16]. Magnetic field is introduced via Peierls substitution. To calculate the LDOS at a single chosen point we used the Kernel Polynomial Expansion algorithms as implemented in Kwant[17]. As a result, we were able to perform calculations for 1μm sized TLG sample with a smooth tip potential of 50 nm size in few minutes on an ordinary laptop.

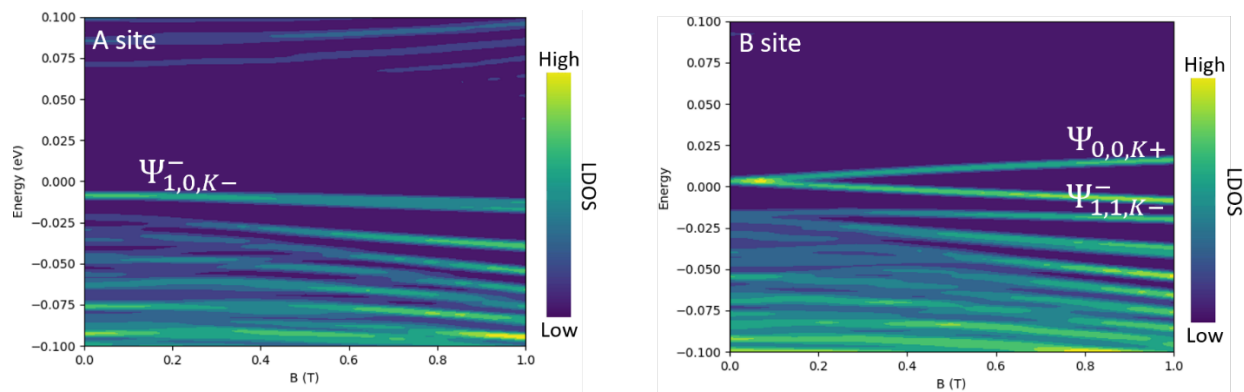

**Figure S9: Calculated local density of states at r=0 for gapped MLG with tip potential.** Example of LDOS for **B** and **A** sublattices in the center of Gaussian potential $U(r) = 50\, e^{-r^2/(2\,(30\text{ nm})^2)}$ meV for gapped MLG with gap $\Delta = 30$ meV. The levels trapped by the potential are marked.

### S8. Localized state due to tip induced potential well

In out experiments, the valence-band monolayer states are pulled up by the p-doping tip potential. As a result, these states are surround by the gap of the monolayer graphene band and will form localized states, see Fig. S10a. Such states will manifest as a $dI/dV_S$ peak in the tunneling spectra. A comparison between the calculated LDOS for sublattice $A_1$ with and without a tip induced potential well at $V_G = 30$ V is shown in Fig. S10b, a strong LDOS peak appeared in the calculation that includes the tip potential well. This explanation is also consistent with the fact that the height of the $dI/dV_S$ peak on sublattice $A_1$ grows with increasing gate voltage in experiment, which, as we show, corresponds to increasing MLG gap size for the entire region and, hence, stronger localization under the tip.

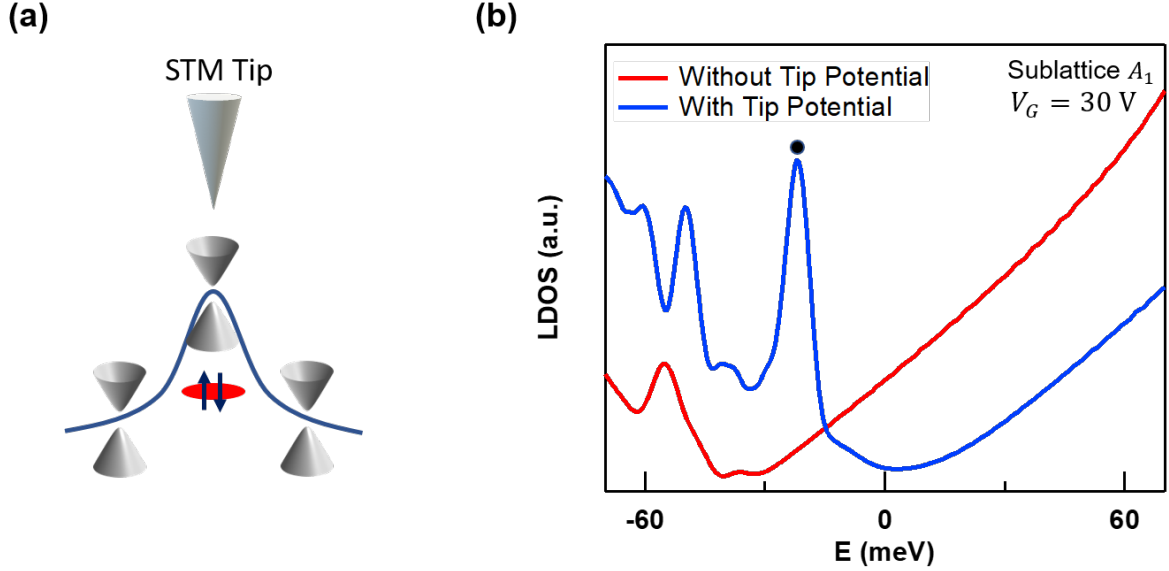

**Figure S10: Localized state due to tip induced potential well. a,** Schematic of the STM tip included potential well (blue line) and the localized stated (red ellipse) from the confinement of the gapped monolayer graphene band. **b,** Calculated tight-binding LDOS for sublattice $A_1$ with and without a tip potential at $V_G = 30$ V. The black dot indicates the state that manifests valley splitting in $B$. The depth and width of the Gaussian potential well used in the simulation are 45 meV and 60 nm, respectively.

## S9. The influence of the tip potential well on the experimental detection of valley splitting in TLG

The presence of the tip induced potential well is important for the experimental detection of the valley splitting that arises from the topological magnetic moment in ABA TLG and its associated large g-factor. Theoretical calculations indicate that a large g-factor with the same order of magnitude as measured in our experiment can be seen from the splitting of the 0-th and -1st Landau level (LL) in the absence of a confinement potential well. However, in small $B$ fields the LL LDOS has weak intensity, giving rise to low signal to noise ratio and complicating experimental detection. But with a confinement potential, stronger localization of the lowest LLs can be achieved, making the corresponding peaks stronger and enhancing the signal to noise ratio; thus, enabling detection of a large g-factor. A comparison between the calculated $LDOS(E, B)$ for sublattice $A_1$ is shown in Fig. S11, the peak splitting, which corresponds to the valley splitting, is only visible with the incorporation of a tip potential well.

In addition, the QD confinement further separates the $0^{th}$ and $-1^{st}$ LLs from the higher LLs due to the stronger influence of the tip potential on the lowest LLs. This separation facilitates detection of individual states because the separation between states is greater than the resolution constraints of our spectroscopic measurement. It also reduces the number of states within an energy range that is equivalent to the resolution constraint of our measurements.

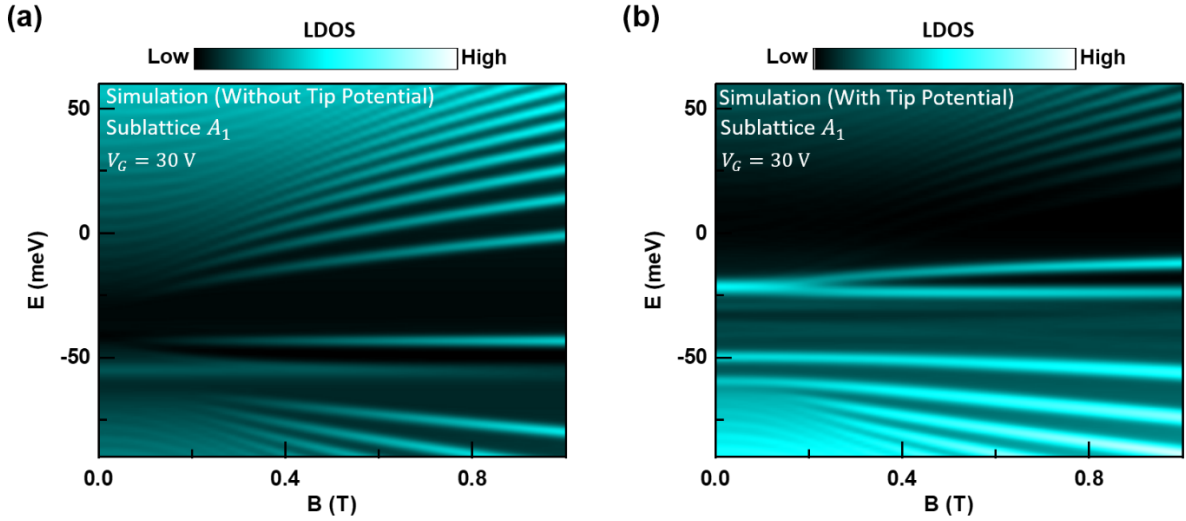

**Figure S11: Enhanced valley splitting visibility with a tip potential. a,** Calculated $LDOS(E, B)$ for sublattice $A_1$ without a tip potential well. No clear peak splitting is visible. **b,** Calculated $LDOS(E, B)$ for sublattice $A_1$ with a tip potential well. A clear peak splitting is visible at around $-25\ meV$. The depth and width of the Gaussian potential well used in the simulation are 45 meV and 60 nm, respectively.

### S10. Tunneling spectra on sublattice $A_2$

In the experiment, we also measured the tunneling spectra on sublattice $A_2$ in zero $B$, the results at several selected gate voltages are shown in Fig. S12. In general, the tunneling spectra on the sublattice $A_2$ location is roughly a mixture of the tunneling spectra on $A_1$ and $B_1$ sublattices. For example, the $dI/dV_S$ intensity at sublattice $A_2$ is always in between the $dI/dV_S$ intensities of sublattices $A_1$ and $B_1$. This result is expected, because in multilayer graphene the layer distance between each graphene layer is ~ 3.4 Å, which is of the same order as the distance between the STM tip and the TLG top layer ($\lesssim$ 5Å). Since the resistance between the tip and sample is set to 60 MΩ in our STS measurements, the resistance between the tip and second layer graphene can be roughly estimated as $60^{\frac{5+3.4}{5}}$ MΩ~1GΩ. Thus, the current contributed from the second layer can be ignored because it is $\lesssim$6% of the tunneling current from the first layer.

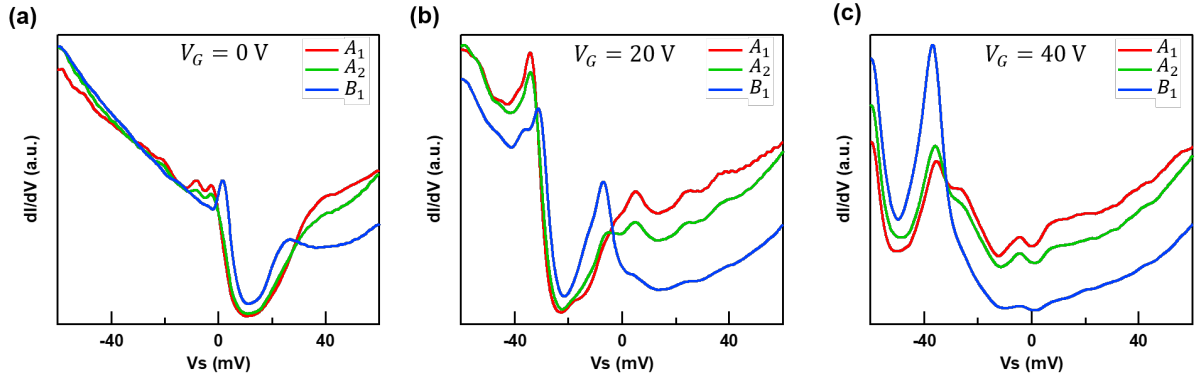

**Figure S12: Tunneling spectra at sublattice $A_1$, $B_1$ and $A_2$ locations. a-c,** Tunneling spectra measured at different sublattice locations at $V_G = 0$ V (a), $V_G = 20$ V (b) and $V_G = 40$ V (c). The set point used to acquire the tunneling spectra in (**a-c**) was $I = 1$ nA, $V_S = -60$ mV, with a 2 mV ac modulation. The tunneling spectra at the $A_2$ sublattice is roughly an average of the $A_1$ and $B_1$ sublattices, due to the greater distance between the $A_2$ sublattice and the STM tip.

**S11. Nonlinear valley splitting in ABA TLG**

The valley splitting in ABA TLG with the STM tip induced QD is complicated because of the competition between the QD confinement and magnetic field confinement, and the mixing of the effective MLG and BLG bands in vertical electric field. These effects lead to a nonlinear valley splitting in ABA TLG as we observed at various $V_G$ in our experiments, see Fig. S13b. The valley splitting ($\Delta E$) are extracted by performing Lorentzian fitting to the split $dI/dV_S$ peaks, a typical fitting result can be seen in Fig. S13a.

First, we discuss the effect of magnetic field confinement on the valley splitting in ABA TLG. In our experiments, we cannot directly measure the potential profile of the tip induced QD, but we expect the size of the QD is relatively large ($\gtrsim 50$ nm). Since the magnetic length $\left(l_B = \sqrt{\frac{\hbar}{eB}}\right)$ in $B = 0.3$ T reaches $\sim 47$ nm, which is already comparable with the QD size, the magnetic confinement will become more important than the QD confinement in larger $B$ and the valley splitting starts to follow the splitting between the 0$^{th}$ and -1$^{st}$ MLG Landau levels $\left(\Delta E \approx \sqrt{\left(\frac{\Delta}{2}\right)^2 + 2\hbar e B v_F^2} - \frac{\Delta}{2} - \delta E\left(\frac{1}{l_B\sqrt{2}}\right)\right)$, which is not linearly dependent on $B$. The last term, $\delta E\left(\frac{1}{l_B\sqrt{2}}\right)$ derived in eq.(8), accounts for the effect of MLG and BLG band mixing on the valley splitting in ABA TLG, see Fig. S14. At even larger magnetic fields that are not reached in our experiments, the -1$^{st}$ MLG Landau level will reach the point of strong mixing with BLG band leading to drastic reduction in the growth of the level-splitting. After considering all of these effects, the theoretical $\Delta E$ achieved is in relatively good agreement with the experiment at all $V_G$, see Fig. S13b. We notice at $V_G = 10$ V the theoretical $\Delta E$ deviates quickly away from the experimental value at high $B$, this is because at $V_G = 10$ V the lower energy valley splitting peak touches the BLG state in high $B$, this makes the MLG and BLG states degenerate, thus the non-degenerate perturbation theory used to calculate the BLG band mixing effect here is not valid anymore (the correction diverges) in this regime.

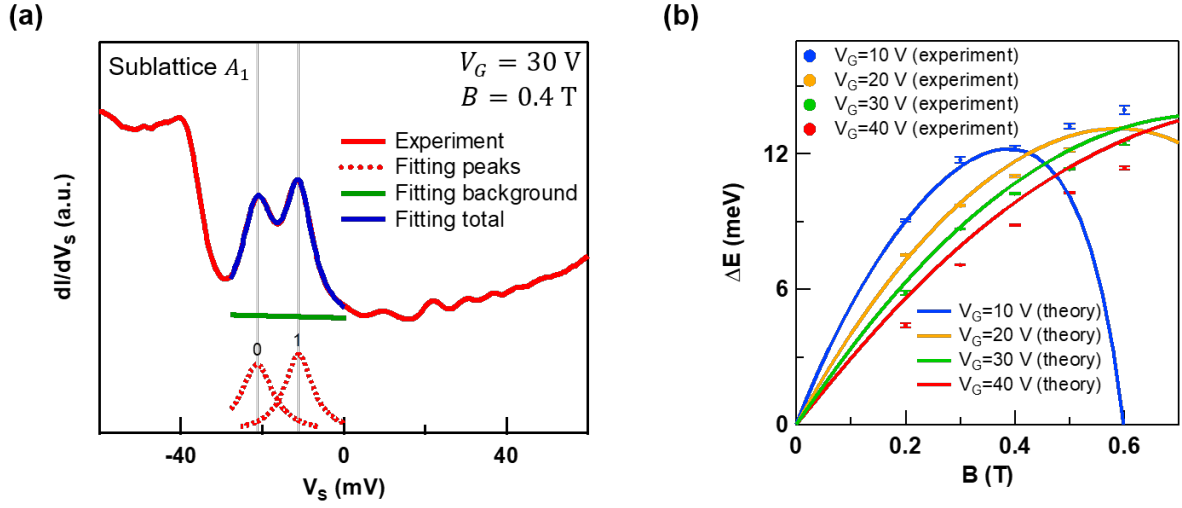

**Figure S13: Nonlinear valley splitting in $B$ at different $V_G$. a,** A typical Lorentzian fitting that we used to extract the $dI/dV_S$ peak positions. **b,** Comparison between the experimental and theoretical valley splitting energy at various $V_G$. The experimental splitting energy is extracted from Fig. S5a-d. The depth and width of the Gaussian potential well used in the theoretical calculation are 50 meV and 40 nm, respectively.

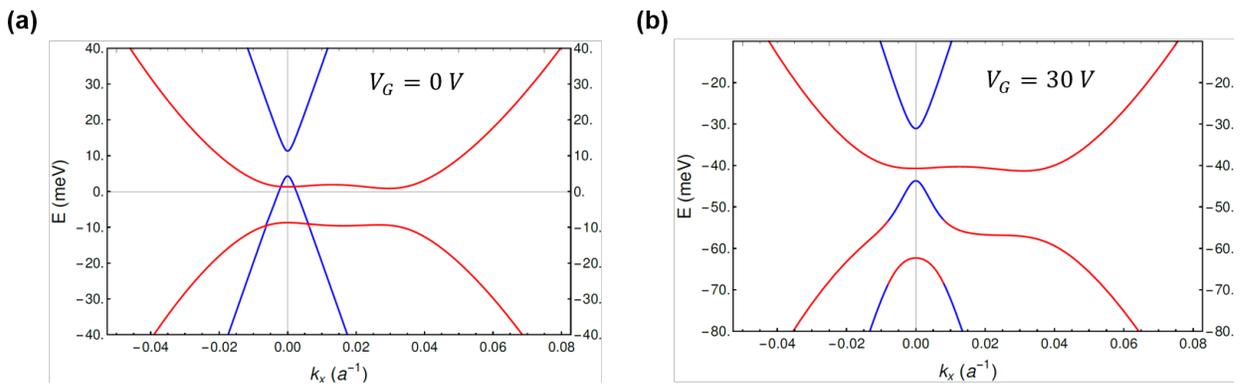

**Figure S14: Electric field induced band mixing. a,** Calculated ABA TLG band structure at $V_G = 0$ V. **b,** Calculated ABA TLG band structure at $V_G = 30$ V. The blue lines correspond to the effective MLG bands, and the red lines correspond to the effective BLG bands. At $V_G = 30$ V, the valence bands of the BLG and MLG parts mix with each other.

## S12. Valley splitting of the effective BLG in ABA TLG

With a simple calculation neglecting the trigonal warping, one can find that the BLG orbital magnetic moment is suppressed by a factor of $\frac{\Delta}{\gamma_1} \approx 1/20$ relative to a monolayer branch if both subsystems have comparable gaps. Therefore, we expect a BLG g-factor of the order of 10-20 in trilayer graphene, which is much smaller than the MLG g-factor. One of the bilayer branches is clearly seen at approximately $-50$ meV in Figs. S18c-d. It is seen on both $A_1$ and $B_1$ sublattices due to gate- and tip- induced hybridization with monolayer branch. There is no obvious splitting observed. The other bilayer branch is localized on the second layer, which cannot be measured by the tunneling spectroscopy in our experiments. Instead, we calculated the LDOS of sublattice $A_2$, which consists solely of the bilayer graphene branch, see Fig. S15b This calculation shows an absence of a strong splitting feature as seen on sublattice $A_1$ (Fig. S15a). The absence of a strong valley splitting of the effective BLG compared to the effective MLG is consistent with the discussion that effective BLG has smaller orbital magnetic moment than effective MLG.

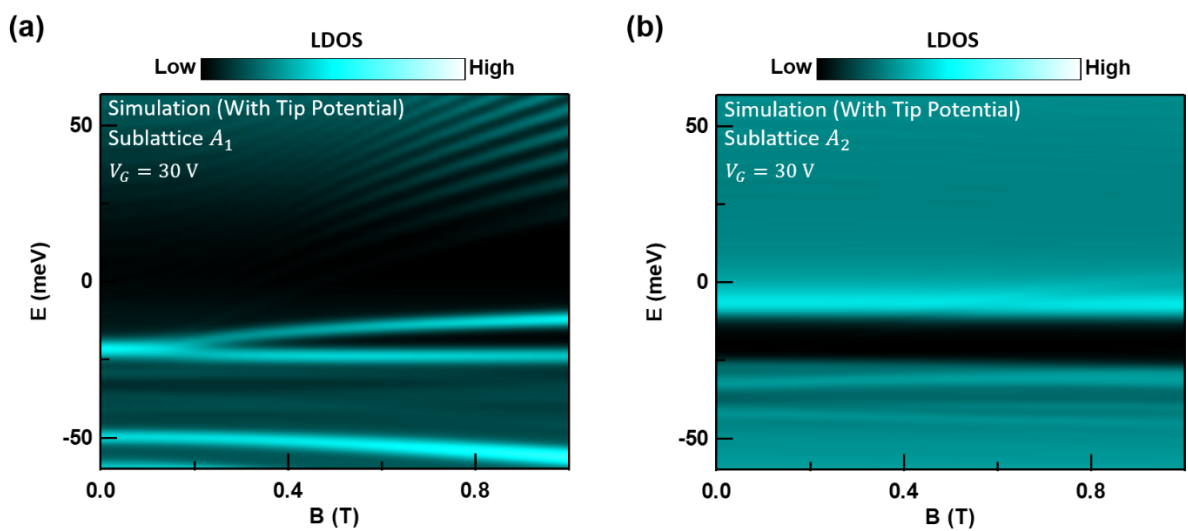

**Figure S15: Calculated $LDOS(E, B)$ for sublattice $A_1$ and $A_2$ at $V_G = 30$ V with a tip potential. a**, Calculated $LDOS(E, B)$ for sublattice $A_1$, where the effective MLG band is localized, a clear valley splitting can be observed at around $E = -25$ meV. **b,** Calculated $LDOS(E, B)$ for sublattice $A_2$, where the effective BLG band localizes, no clear valley splitting is seen, only a faint valley splitting is visible near $E \sim -48\ meV$. Because this splitting is faint and the $A_2$ sublattice is located in the middle TLG layer, experimental detection of this splitting is unlikely. The depth

and width of the Gaussian potential well used in the simulations are 45 meV and 60 nm, respectively.

## S13. Linear fit to the valley splitting in small $B$ field

The upper panels of Figs. S16a-d show the measured $dI^3/dV_S^3(V_S, B)$ at different $V_G$ with the same STM tip and at the same region as in Fig. 4, the valley splitting state and Landau level fans can both be observed in these measurements. The lower panels of Figs. S16a-d show the zoom-in of the valley splitting state at different $V_G$ from the upper panel of Figs. S16a-d, the valley splitting is not linearly dependent on $B$ within the full experimentally measured $B$ range but approximately linearly dependent on $B$ in small $B$ region. We extracted the $V_S$ value for the split peak at different $B$ and $V_G$ by doing Lorentzian fitting to the $dI^3/dV_S^3$ data, see Figs. S17b-e. A typical Lorentzian fitting result can be seen in Fig. S17a. Then we performed linear fitting to the extract split peak positions in small $B$ to calculate $g_v$ based on the slope of the linear fitting lines, see Figs. S17b-e. The fitted lines are also plotted as dashed yellow lines in Fig. 16. We extracted $g_v$ with a value of around $1050, 722, 611$ and $517$ for $V_G = 10\,\text{V}, 20\,\text{V}, 30\text{V}$ and $40\,\text{V}$, respectively.

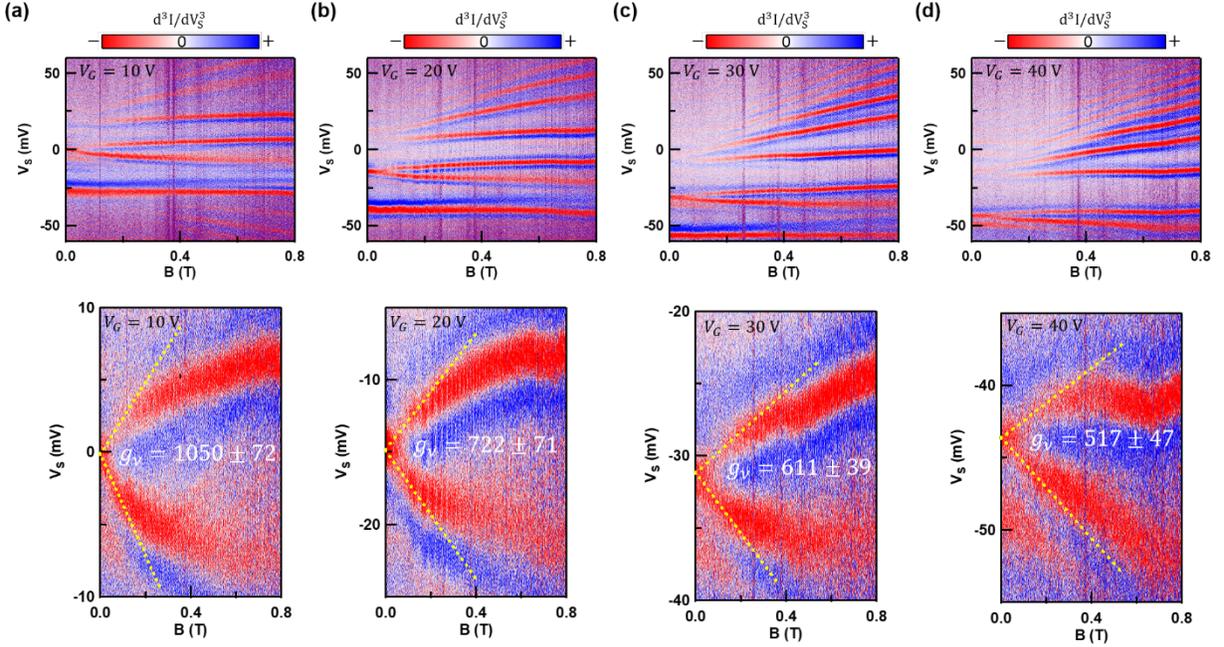

**Figure S16: Valley g-factor extraction at different $V_G$. a-d,** Measured $d^3I/dV_S^3$ $(V_S, B)$ at $V_G = 10$ V (a), $V_G = 20$ V (b), $V_G = 30$ V (c) and $V_G = 40$ V (d). The lower panels in a-d are a zoom in of the data shown in the upper panel of a-d, respectively. The $d^3I/dV_S^3$ values are numerically calculated from the $dI/dV_S$ data measured from the lock-in in experiment. The purpose of showing $d^3I/dV_S^3$ data is to better visualize the peak splitting. The red features in $d^3I/dV_S^3$ correspond to the peak positions in $dI/dV_S$ spectra. The yellow dashed lines in the lower panel of a-d are the linear fit to the peak splitting in small $B$. The valley g-factor is calculated based on the slope of the linear fits. The set point used to acquire the $dI/dV_S$ tunneling spectra in **(a-d)** was $I = 1$ nA, $V_S = -60$ mV, with a 2 mV ac modulation. The STM tip used to acquire the data in **(a-d)** does not have good atomic resolution for the tunneling spectra.

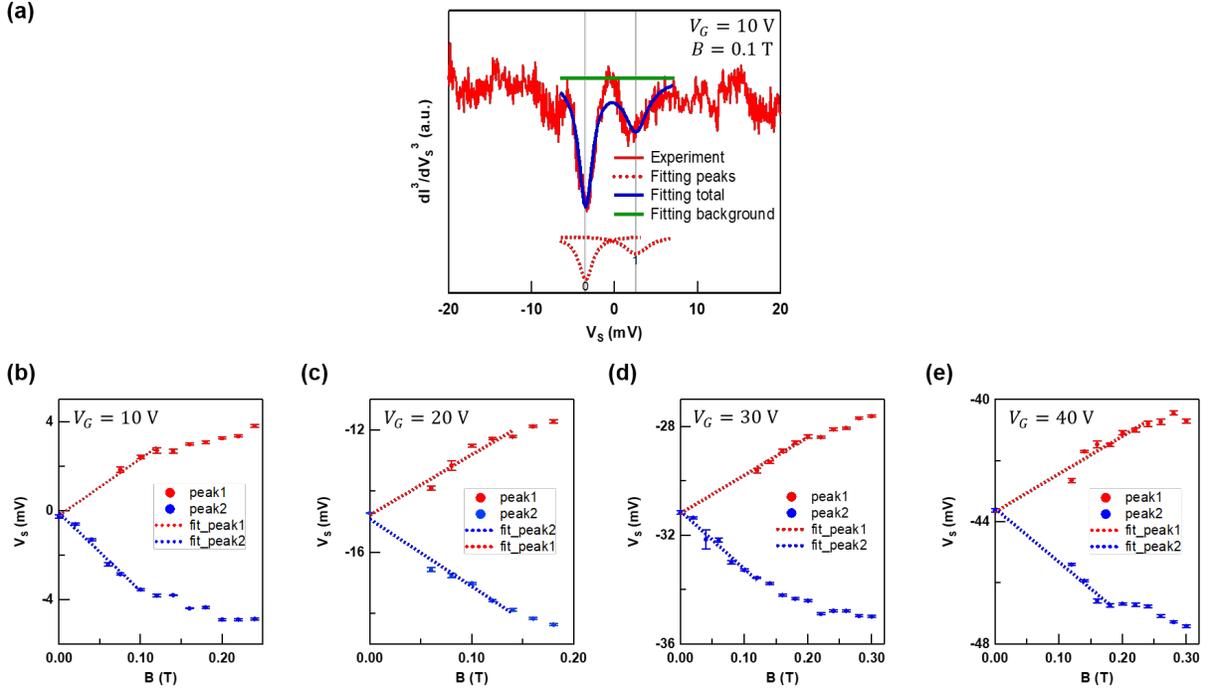

**Figure S17: Peak position extraction and linear fitting of peaks positions in small $B$. a,** a typical Lorentzian fitting for the experimental $dI^3/dV_S^3$ data that we use to extract the peak positions of the valley splitting peak. **b-e,** Extracted peak positions at different $B$ from the $dI^3/dV_S^3$ data in Fig. S16 and the linear fitting to the peak positions in small $B$ at $V_G = $ 10 V, 20 V, 30 V and 40 V, respectively.

## S14. Sublattice resolved $dI/dV_S(V_S, B)$ color plots

The sublattice resolved simulated $LDOS(E, B)$ and measured $dI/dV_S(V_S, B)$ at $V_G = 30$ V are compared, they have good agreement. On sublattice $A_1$, the split peaks associated with valley splitting appeared in both simulation (Fig. S18a) and experiment (Fig. S18c), and the valley splitting energy is in good agreement between them. Whereas on sublattice $B_1$, the split peaks are absent in both simulation (Fig. S18b) and experiment (Fig. S18d). In addition, both simulations and experiments show Landau level fans on sublattice $A_1$ and sublattice $B_1$ with nonuniform energy spacing. Finally, we noticed there is a discrepancy between the simulation and experiment for the LDOS on sublattice $B_1$, for which simulation predict a stronger peak at $E \approx -25$ meV than observed in the experiment.

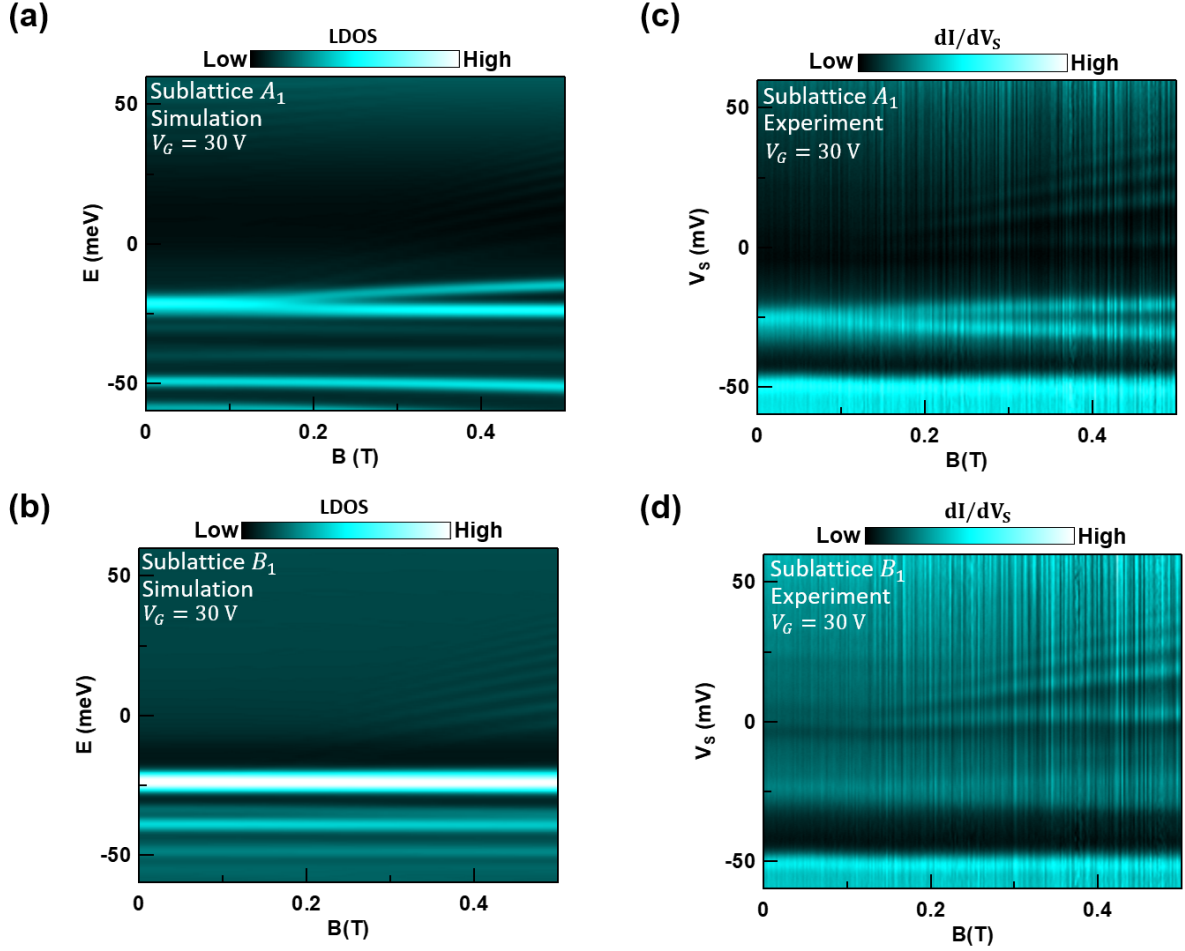

**Figure S18: Sublattice resolved experimentally measured $dI/dV_S(V_S, B)$ and simulated $LDOS(E, B)$. a-b,** Simulated $LDOS(E, B)$ color plots on sublattice $A_1$ **(a)** and $B_1$ **(b)** for an ABA TLG QD at $V_G = 30$ V. The depth and width of the Gaussian potential well used in the simulation are 45 meV and 60 nm, respectively. **c-d,** Experimentally measured $dI/dV_S(V_S, B)$ color plots on sublattice $A_1$ **(c)** and $B_1$ **(d)** at $V_G = 30$ V. The tunneling spectra were measured with a different calibrated STM tip and at a different location on the sample compared to the data presented in Figs. 3 and 4. The set point used to acquire the tunneling spectra was $I = 1$ nA, $V_S = -60$ mV, with a 2 mV ac modulation.